\documentclass[9pt,twocolumn,twoside]{IEEEtran} 
\usepackage{amssymb}
\usepackage{graphicx,subfigure}
\usepackage[centertags]{amsmath}
\usepackage{epsfig,algorithm,amsmath,amssymb,color,subfigure,url}
\newtheorem{theorem}{Theorem}
\newtheorem{lemma}{Lemma}

\newtheorem{corollary}{Corollary}
\newtheorem{definition}{Definition}
\long\def\comment#1{}

\begin{document}
\renewcommand{\textfraction}{0}

\title{KF-CS: Compressive Sensing on Kalman Filtered Residual}

\author{Namrata Vaswani
\thanks{N. Vaswani is with the ECE dept at Iowa State University, Ames, IA (email: namrata@iastate.edu).
This research was partially supported by NSF grants ECCS-0725849 and CCF-0917015. A part of this work appeared in \cite{kfcsicip, kfcspap}.
}
}
\date{}
\maketitle \pagestyle{plain} 

\setlength{\arraycolsep}{0.03cm}
\newcommand{\xhat}{\hat{x}}
\newcommand{\xpred}{\hat{x}_{t|t-1}}
\newcommand{\Ppred}{P_{t|t-1}}
\newcommand{\ty}{\tilde{y}_t}
\newcommand{\tty}{\tilde{y}_{t,\text{res}}}
\newcommand{\tw}{\tilde{w}_t}
\newcommand{\ttw}{\tilde{w}_{t,f}}
\newcommand{\betahat}{\hat{\beta}}

\newcommand{\ypast}{y_{1:t-1}}
\newcommand{\sone}{S_{*}}
\newcommand{\sinf}{{S_{**}}}
\newcommand{\smax}{S_{\max}}
\newcommand{\smin}{S_{\min}}
\newcommand{\samax}{S_{a,\max}}
\newcommand{\Nhat}{{\hat{N}}}

\newcommand{\Dnum}{D_{num}}
\newcommand{\pss}{p^{**,i}}
\newcommand{\fr}{f_{r}^i}

\newcommand{\A}{{\cal A}}
\newcommand{\Z}{{\cal Z}}
\newcommand{\B}{{\cal B}}
\newcommand{\R}{{\cal R}}
\newcommand{\reg}{{\cal G}}
\newcommand{\const}{\mbox{const}}

\newcommand{\trace}{\mbox{tr}}

\newcommand{\hsim}{{\hspace{0.0cm} \sim  \hspace{0.0cm}}}
\newcommand{\he}{{\hspace{0.0cm} =  \hspace{0.0cm}}}

\newcommand{\vect}[2]{\left[\begin{array}{cccccc}
     #1 \\
     #2
   \end{array}
  \right]
  }

\newcommand{\matr}[2]{ \left[\begin{array}{cc}
     #1 \\
     #2
   \end{array}
  \right]
  }
\newcommand{\vc}[2]{\left[\begin{array}{c}
     #1 \\
     #2
   \end{array}
  \right]
  }

\newcommand{\gdot}{\dot{g}}
\newcommand{\Cdot}{\dot{C}}
\newcommand{\re}{\mathbb{R}}
\newcommand{\n}{{\cal N}}  
\newcommand{\N}{{\overrightarrow{\bf N}}}  
\newcommand{\chat}{\tilde{C}_t}
\newcommand{\chati}{\chat^i}

\newcommand{\cmin}{C^*_{min}}
\newcommand{\twi}{\tilde{w}_t^{(i)}}
\newcommand{\twj}{\tilde{w}_t^{(j)}}
\newcommand{\wi}{{w}_t^{(i)}}
\newcommand{\twio}{\tilde{w}_{t-1}^{(i)}}

\newcommand{\tWi}{\tilde{W}_n^{(m)}}
\newcommand{\tWj}{\tilde{W}_n^{(k)}}
\newcommand{\Wi}{{W}_n^{(m)}}
\newcommand{\tWio}{\tilde{W}_{n-1}^{(m)}}

\newcommand{\ds}{\displaystyle}

\newcommand{\SAR}{S$\!$A$\!$R }
\newcommand{\MAR}{MAR}
\newcommand{\MMRF}{MMRF}
\newcommand{\AR}{A$\!$R }
\newcommand{\GMRF}{G$\!$M$\!$R$\!$F }
\newcommand{\DTM}{D$\!$T$\!$M }
\newcommand{\MSE}{M$\!$S$\!$E }
\newcommand{\RCS}{R$\!$C$\!$S }
\newcommand{\uomega}{\underline{\omega}}
\newcommand{\y}{v}
\newcommand{\x}{w}
\newcommand{\lu}{\mu}
\newcommand{\g}{g}
\newcommand{\s}{{\bf s}}
\newcommand{\bft}{{\bf t}}
\newcommand{\refmap}{{\cal R}}
\newcommand{\totrefl}{{\cal E}}
\newcommand{\beq}{\begin{equation}}
\newcommand{\eeq}{\end{equation}}
\newcommand{\bdm}{\begin{displaymath}}
\newcommand{\edm}{\end{displaymath}}
\newcommand{\hatz}{\hat{z}}
\newcommand{\hatu}{\hat{u}}
\newcommand{\tilz}{\tilde{z}}
\newcommand{\tilu}{\tilde{u}}
\newcommand{\hhatz}{\hat{\hat{z}}}
\newcommand{\hhatu}{\hat{\hat{u}}}
\newcommand{\tilc}{\tilde{C}}
\newcommand{\hatc}{\hat{C}}
\newcommand{\tim}{n}

\newcommand{\ssp}{\renewcommand{\baselinestretch}{1.0}}
\newcommand{\defd}{\mbox{$\stackrel{\mbox{$\triangle$}}{=}$}}
\newcommand{\goes}{\rightarrow}
\newcommand{\tends}{\rightarrow}
\newcommand{\defn}{\triangleq} 
\newcommand{\se}{&=&}
\newcommand{\sdefn}{& \defn  &}
\newcommand{\sle}{& \le &}
\newcommand{\sge}{& \ge &}
\newcommand{\plusminus}{\stackrel{+}{-}}
\newcommand{\Ey}{E_{Y_{1:t}}}
\newcommand{\ey}{E_{Y_{1:t}}}

\newcommand{\equivto}{\mbox{~~~which is equivalent to~~~}}
\newcommand{\nonzero}{i:\pi^n(x^{(i)})>0}
\newcommand{\nonzeroc}{i:c(x^{(i)})>0}

\newcommand{\supn}{\sup_{\phi:||\phi||_\infty \le 1}}
\newtheorem{remark}{Remark}
\newtheorem{example}{Example}
\newtheorem{ass}{Assumption}
\newtheorem{proposition}{Proposition}

\newtheorem{fact}{Fact}
\newtheorem{heuristic}{Heuristic}
\newcommand{\eps}{\epsilon}
\newcommand{\bd}{\begin{definition}}
\newcommand{\ed}{\end{definition}}
\newcommand{\udq}{\underline{D_Q}}
\newcommand{\td}{\tilde{D}}
\newcommand{\epsinv}{\epsilon_{inv}}
\newcommand{\al}{\mathcal{A}}

\newcommand{\bfx} {\bf X}
\newcommand{\bfy} {\bf Y}
\newcommand{\bfz} {\bf Z}
\newcommand{\ddas}{\mbox{${d_1}^2({\bf X})$}}
\newcommand{\ddbs}{\mbox{${d_2}^2({\bfx})$}}
\newcommand{\dda}{\mbox{$d_1(\bfx)$}}
\newcommand{\ddb}{\mbox{$d_2(\bfx)$}}
\newcommand{\xinc}{{\bfx} \in \mbox{$C_1$}}
\newcommand{\eqa}{\stackrel{(a)}{=}}
\newcommand{\eqb}{\stackrel{(b)}{=}}
\newcommand{\eqe}{\stackrel{(e)}{=}}
\newcommand{\leqc}{\stackrel{(c)}{\le}}
\newcommand{\leqd}{\stackrel{(d)}{\le}}

\newcommand{\leqa}{\stackrel{(a)}{\le}}
\newcommand{\leqb}{\stackrel{(b)}{\le}}
\newcommand{\leqe}{\stackrel{(e)}{\le}}
\newcommand{\leqf}{\stackrel{(f)}{\le}}
\newcommand{\leqg}{\stackrel{(g)}{\le}}
\newcommand{\leqh}{\stackrel{(h)}{\le}}
\newcommand{\leqi}{\stackrel{(i)}{\le}}
\newcommand{\leqj}{\stackrel{(j)}{\le}}

\newcommand{\w}{{W^{LDA}}}
\newcommand{\halpha}{\hat{\alpha}}
\newcommand{\hsigma}{\hat{\sigma}}
\newcommand{\slmax}{\sqrt{\lambda_{max}}}
\newcommand{\slmin}{\sqrt{\lambda_{min}}}
\newcommand{\lmax}{\lambda_{max}}
\newcommand{\lmin}{\lambda_{min}}

\newcommand{\da} {\frac{\alpha}{\sigma}}
\newcommand{\chka} {\frac{\check{\alpha}}{\check{\sigma}}}
\newcommand{\sumo}{\sum _{\underline{\omega} \in \Omega}}
\newcommand{\distance}{d\{(\hatz _x, \hatz _y),(\tilz _x, \tilz _y)\}}
\newcommand{\col}{{\rm col}}
\newcommand{\rcs}{\sigma_0}
\newcommand{\CalR}{{\cal R}}
\newcommand{\df}{{\delta p}}
\newcommand{\dq}{{\delta q}}
\newcommand{\dZ}{{\delta Z}}
\newcommand{\pprime}{{\prime\prime}}

\newcommand{\vn}{N}

\newcommand{\bv}{\begin{vugraph}}
\newcommand{\ev}{\end{vugraph}}
\newcommand{\bi}{\begin{itemize}}
\newcommand{\ei}{\end{itemize}}
\newcommand{\ben}{\begin{enumerate}}
\newcommand{\een}{\end{enumerate}}
\newcommand{\be}{\protect\[}
\newcommand{\ee}{\protect\]}
\newcommand{\bean}{\begin{eqnarray*} }
\newcommand{\eean}{\end{eqnarray*} }
\newcommand{\bea}{\begin{eqnarray} }
\newcommand{\eea}{\end{eqnarray} }
\newcommand{\nn}{\nonumber}
\newcommand{\ba}{\begin{array} }
\newcommand{\ea}{\end{array} }
\newcommand{\ep}{\mbox{\boldmath $\epsilon$}}
\newcommand{\epp}{\mbox{\boldmath $\epsilon '$}}
\newcommand{\Lep}{\mbox{\LARGE $\epsilon_2$}}
\newcommand{\und}{\underline}
\newcommand{\pdif}[2]{\frac{\partial #1}{\partial #2}}
\newcommand{\odif}[2]{\frac{d #1}{d #2}}
\newcommand{\dt}[1]{\pdif{#1}{t}}
\newcommand{\urho}{\underline{\rho}}

\newcommand{\spc}{{\cal S}}
\newcommand{\tspc}{{\cal TS}}

\newcommand{\uv}{\underline{v}}
\newcommand{\us}{\underline{s}}
\newcommand{\uc}{\underline{c}}
\newcommand{\utheta}{\underline{\theta}^*}
\newcommand{\ualpha}{\underline{\alpha^*}}

\newcommand{\uxy}{\underline{x}^*}
\newcommand{\uxyj}{[x^{*}_j,y^{*}_j]}
\newcommand{\arcl}[1]{arclen(#1)}
\newcommand{\one}{{\mathbf{1}}}

\newcommand{\uxyjt}{\uxy_{j,t}}
\newcommand{\E}{\mathbb{E}}

\newcommand{\rhomat}{\left[\begin{array}{c}
                        \rho_3 \ \rho_4 \\
                        \rho_5 \ \rho_6
                        \end{array}
                   \right]}
\newcommand{\deltat}{\tau} 
\newcommand{\deltatt}{\Delta t_1}
\newcommand{\ceil}[1]{\ulcorner #1 \urcorner}

\newcommand{\xxi}{x^{(i)}}
\newcommand{\txi}{\tilde{x}^{(i)}}
\newcommand{\txj}{\tilde{x}^{(j)}}

\newcommand{\mi}[1]{{#1}^{(m,i)}}

\setlength{\arraycolsep}{0.05cm}

\newcommand{\Section}[1]{ \vspace{-0.075in} \section{#1} \vspace{-0.025in} } 
\newcommand{\Subsection}[1]{\vspace{-0.075in}  \subsection{#1}  \vspace{-0.02in} } 
\newcommand{\Subsubsection}[1]{   \subsubsection{#1} } 
\newcommand{\rest}{{T_\text{rest}}}
\newcommand{\zetahat}{\hat{\zeta}}
\newcommand{\sm}{\text{small}}
\newcommand{\tDelta}{{\tilde{\Delta}}}
\newcommand{\tDeltae}{{\tilde{\Delta}_e}}
\newcommand{\tT}{{\tilde{T}}}
\newcommand{\add}{{\cal A}}
\newcommand{\rem}{{\cal R}}
\newtheorem{sigmodel}{Signal Model}
\newcommand{\thr}{{\text{thr}}}
\newcommand{\delthr}{{\text{del-thr}}}
\newcommand{\delbound}{{b}}
\newcommand{\err}{{\text{err}}}
\newcommand{\Q}{{\cal Q}}
\newcommand{\dett}{{\text{det}}}
\newcommand{\CSres}{{\text{CSres}}}
\newcommand{\diff}{{\text{diff}}}
\newcommand{\ssim}{& \sim & }


\begin{abstract}
We consider the problem of recursively reconstructing time sequences of sparse signals (with unknown and time-varying sparsity patterns) from a limited number of linear incoherent measurements with additive noise. The idea of our proposed solution, KF CS-residual (KF-CS) is to replace compressed sensing (CS) on the observation by CS on the Kalman filtered (KF) observation residual computed using the previous estimate of the support. KF-CS error stability over time is studied. Simulation comparisons with CS and LS-CS are shown.
\end{abstract}

\Section{Introduction} 
\label{intro}
Consider the problem of recursively and causally reconstructing time sequences of spatially  sparse signals (with unknown and time-varying sparsity patterns) from a limited number of linear incoherent measurements with additive noise. The signals are sparse in some transform domain referred to as the sparsity basis. Important applications include dynamic MRI reconstruction for real-time applications such as MRI-guided surgery, single-pixel video imaging \cite{singlepixelvideo}, or video compression.
%
Due to strong temporal dependencies in the signal sequence, it is usually valid to assume that its {\em sparsity pattern (support of the sparsity transform vector) changes slowly over time}. This was verified in \cite{just_lscs,kfcsmri}. 

The solution to the static version of the above problem is provided by compressed sensing (CS) \cite{candes,donoho}. CS for noisy observations, e.g. Dantzig selector  \cite{dantzig}, Lasso \cite{candes_rip}, or Basis Pursuit Denoising (BPDN) \cite{bpdn,tropp}, have been shown to have small error as long as incoherence assumptions hold.  Most existing solutions for the dynamic problem, e.g. \cite{singlepixelvideo,sparsedynamicMRI}, are non-causal and batch solutions. Batch solutions process the entire time sequence in one go and thus have much higher reconstruction complexity. An alternative would be to apply CS at each time separately (simple CS), which is online and low-complexity, but since it does not use past observations, its reconstruction error is much larger when the number of available observations is small.
{\em Our goal} is to develop a recursive solution that improves the accuracy of simple CS by using past observations, but keeps the reconstruction complexity similar to that of simple CS. By {\em ``recursive", we mean a solution that uses only the previous signal estimate and the current observation vector at the current time.}

In this work, we propose a solution called KF-CS-residual (KF-CS) which is motivated by reformulating the above problem as causal minimum mean squared error (MMSE) estimation with a slow time-varying set of dominant basis directions (or equivalently the support of the sparsity basis coefficients' vector). If the support is known, and a linear Gaussian prior dynamic model is assumed for the nonzero coefficients, the causal MMSE solution is given by the Kalman filter (KF) \cite{ekf} for this support. When the support is unknown and time-varying, the initial support can be estimated using CS. Whenever there is an addition to the support, it can be estimated by running CS on the KF residual, followed by thresholding. This new support estimate can be used to run the KF at the next time instant. If some coefficients become and remain nearly zero, they can be removed from the support set. Both the computational and storage complexity of KF-CS is similar to that of simple CS  - $O(m^3)$ at a given time where $m$ is the signal length  \cite[Table 1]{ldpc_cs} and $O(Nm^3)$ for an $N$ length sequence. This is significantly lower than  $O(N^3m^3)$ for batch CS. 
%
Note that a full KF, that does not use the knowledge that the signal is sparse, is meaningless here, because the number of observations
available is smaller than the signal dimension, and thus many elements of the state (sparsity basis coefficients vector) will be unobservable. Unless all unobservable modes are stable, the error will blow up \cite{ekf,kfcsicip}.


The most closely related work to KF-CS is our work on LS-CS \cite{kfcspap,just_lscs} which uses an LS residual instead of a KF residual. Thus it only uses the previous support estimate, not the previous signal estimates, to improve the current reconstruction. KF-CS uses both and hence it outperforms LS-CS when the available number of measurements is small, e.g. see Fig. \ref{lscs_kfcs}. The work of \cite{jung_etal} gives an approximate batch-CS approach for dynamic MRI. Bayesian approaches, but all for reconstructing a single sparse signal, include \cite{bayesianCS,schniter,ldpc_cs}. Related work, which appeared after \cite{kfcsicip}, and in parallel with \cite{kfcspap}, includes \cite{giannakis} (addresses recursive sparse estimation but with time-invariant support), and our own later work on modified-CS \cite{isitmodcs}.




{\em This paper is organized as follows.} The signal model and the algorithm are described in Sec. \ref{priorsignalmodel}. We analyze the CS-residual step of KF-CS in Sec. \ref{whycsresworks}. In Sec. \ref{supplem_kfcsstab}, we prove KF-CS error stability and discuss why our result needs stronger assumptions than a similar result for LS-CS \cite{just_lscs}. Simulation results comparing KF-CS with LS-CS and simple CS are given in Sec. \ref{sims} and conclusions in Sec. \ref{conclusions}.%
%
%

In this work, {\em we do ``CS", whether in simple CS or in CS-residual, using the Dantzig selector (DS) \cite{dantzig}}. This choice was initially motivated by the fact that its guarantees are stronger (depend only on signal support size, not support elements) than those for BPDN \cite{tropp} and its results are simpler to apply and modify. In later work \cite{kfcsmri}, we have also used BPDN. Between DS and Lasso \cite{candes_rip}, either can be used and everything will remain the same except for some constants.%

\Subsection{Notation and Problem Definition}
\label{notation}

The set operations $\cup$, $\cap$, and $\setminus$ have the usual meanings. $T^c$ denotes the complement of $T$ w.r.t. $[1,m]:=[1,2,\dots m]$, i.e. $T^c := [1,m] \setminus T$. $|T|$ denotes the size (cardinality) of $T$.

For a vector, $v$, and a set, $T$, $v_T$ denotes the $|T|$ length sub-vector containing the elements of $v$ corresponding to the indices in the set $T$. $\|v\|_k$ denotes the $\ell_k$ norm of a vector $v$. If just $\|v\|$ is used, it refers to $\|v\|_2$.
%
%
For a matrix $M$, $\|M\|_k$ denotes its induced $k$-norm, while just $\|M\|$ refers to $\|M\|_2$. $M'$ denotes the transpose of $M$. For a tall matrix, $M$, $M^\dag : = (M'M)^{-1}M'$. For symmetric matrices, $M_1 \le M_2$ means that $M_2-M_1$ is positive semidefinite.
For a fat matrix $A$, $A_T$ denotes the sub-matrix obtained by extracting the columns of $A$ corresponding to the indices in $T$.
The $S$-restricted isometry property (RIP) constant, $\delta_S$, and the $S,S'$-restricted orthogonality constant, $\theta_{S,S'}$, are as defined in equations 1.3 and 1.5 of \cite{dantzig} respectively.


For a square matrix, $Q$, we use $(Q)_{T_1,T_2}$ to denote the sub-matrix of $Q$ containing rows and columns corresponding to the entries in $T_1$ and $T_2$ respectively. $I$ denotes an appropriate sized identity matrix.
{\em The $m \times m$ matrix $I_T$ is defined as}
\bea
(I_T)_{T,T} = I, \ (I_T)_{T^c,[1,m]} = 0, \ (I_T)_{[1,m], T^c} = 0
\label{defI}
\eea
We use $0$ to denote a vector or matrix of all zeros of appropriate size. The notation $z \sim \n(\mu,\Sigma)$ means that $z$ is Gaussian distributed with mean $\mu$ and covariance $\Sigma$.


Let $(z_t)_{m \times 1}$ denote the spatial signal at time $t$ and $(y_t)_{n \times 1}$, with $n<m$, denote its noise-corrupted observation vector at $t$, i.e. $y_t = H z_t + w_t$. The signal, $z_t$, is sparse in a given sparsity basis (e.g. wavelet) with orthonormal basis matrix, $\Phi_{m \times m}$, i.e. $x_t \defn \Phi' z_t$ is a sparse vector. We denote its support by $N_t$ and we use $S_t:= |N_t|$ to denote its size. 
Thus the observation model is
\bea
y_t = A x_t + w_t, \ A \defn H \Phi, \ \  \E[w_t]=0, \  \E[w_t w_t']= \sigma^2 I \ \ \ \
\label{obsmod}
\eea
where $\E[\cdot]$ denotes expectation. We assume that $A$ has unit norm columns. The observation noise, $w_t$, is independent identically distributed (i.i.d.) over $t$ and is independent of $x_t$. Our goal is to recursively estimate $x_t$ (or equivalently the signal, $z_t = \Phi x_t$) using $y_1, \dots y_t$. By {\em recursively}, we mean, use only $y_t$ and the estimate from $t-1$, $\xhat_{t-1}$, to compute the estimate at $t$.


\bd[Define $\sone$, $\sinf$]
For $A:=H \Phi$, 
\ben
\label{deltas}
\item let $\sone$ denote the largest $S$ for which $\delta_S < 1/2$,
\label{delta3s}
\item let $\sinf$ denote the largest $S$ for which $\delta_{2S}+ \theta_{S,2S} < 1$.
\een
\label{delta_bnd}
\ed

\bd[Define $\xhat_t$, $\Nhat_t$]
We use $\xhat_t$ to denote the final estimate of $x_t$ at time $t$ and $\Nhat_t$ to denote its support estimate.
\ed


\bd[Define $T$, $\Delta$, $\Delta_e$]
We use $T  \equiv T_t := \Nhat_{t-1}$ to denote the support estimate from the previous time. This serves as an initial estimate of the current support.
We use $\Delta \equiv \Delta_t := N_t \setminus T_t$ to denote the unknown part of the support at the current time. 
We use $\Delta_e \equiv \Delta_{e,t} := T_t \setminus N_t$ to denote the ``erroneous" part of $T_t$. To keep notation simple, we remove the subscript $t$ in most places.
\ed

%
%


\Section{Kalman Filtered CS residual (KF-CS)}
\label{priorsignalmodel}

The LS-CS-residual (LS-CS) algorithm \cite{just_lscs} only used the previous support estimate, $T$, to obtain the current reconstruction, but did not use the previous nonzero coefficient estimates, $(\xhat_{t-1})_T$. Because of temporal dependencies, these also change slowly and using this fact should improve reconstruction accuracy further.
To do this we can replace LS by regularized LS. If training data is available to learn a linear prior model for signal coefficients' change, this can be done by replacing the initial LS estimate of LS-CS by a Kalman filtered (KF) estimate. The KF will give the optimal (in terms of minimizing the Bayesian MSE) regularization parameters if the size of the unknown support, $|\Delta|=0$. These will be close-to-optimal if $|\Delta|$ is nonzero but small. We assume a simple linear model described below in Sec. \ref{signalmodel}. We develop the KF-CS algorithm for it in Sec. \ref{kfcsalgo} and discuss its pros and cons in Sec. \ref{whyrandomwalk}.

\Subsection{Signal Model} 
\label{signalmodel}
We assume an i.i.d. Gaussian random walk model with support additions and removals occurring every $d$ time instants. Additions occur at every $t_j=1+jd$ and removals at every $t_{j+1}-1$ for all $j \ge 0$. The support sets, $N_t$, at all $t$, are deterministic unknowns, while the sequence of $x_t$'s is a random process.%
\begin{sigmodel}
Assume the following model.
\ben

\item  At $t=0$, $x_0$ is $S_0$ sparse with support $N_0$ and $(x_0)_{N_0} \sim \n(0,\sigma_{sys,0}^2 I)$.

\item At every addition time, $t_j = 1 + jd$, for all $j \ge 0$, there are $S_a$ new additions to the support. Denote the set of indices of the coefficients added at $t_j$ by $\add(j)$.

\item At every removal time, $t_{j+1}-1 = (j+1)d$, for all $j \ge 0$, there are $S_r$ removals from the support.


\item The maximum support size is $\smax$, i.e. $|N_t| \le \smax$ at all $t$.

\item Every new coefficient that gets added to the support starts from zero and follows an independent Gaussian random walk model with zero drift and change variance $\sigma_{sys}^2$.

\item The value of every removed coefficient and the corresponding change variance both  get set to zero. 

\een
The above model can be summarized as follows.
\bea
|N_t \setminus N_{t-1}| \se
\left\{ \begin{array}{cc} 
S_a \ & \ \text{if} \ t=t_j  \  \  \  \  \ \\
0 \ & \ \text{otherwise}  \  \  \  \  \
\end{array}
\right. \nn \\
|N_{t-1} \setminus N_t| \se
\left\{ \begin{array}{cc} 
S_r \ & \ \text{if} \ t=t_{j+1}-1   \  \  \  \  \ \\
0 \ & \ \text{otherwise}  \  \  \  \  \
\end{array}
\right. \nn \\
x_0 \ssim \n(0,Q_0),  \  \text{where} \  Q_0 = \sigma_{sys,0}^2 I_{N_0}  \nn \\
\nu_t \ssim \n(0,Q_t),   \  \text{where} \  Q_t = \sigma_{sys}^2 I_{N_t} \nn \\
(x_t)_{N_t}  \se   (x_{t-1})_{N_t} + (\nu_t)_{N_t}   \nn \\
(x_t)_{N_t^c} \se (\nu_t)_{N_t^c} =  0
\label{sysmod}
\eea

\label{randomwalk}
\end{sigmodel}
\begin{ass} We assume that
\ben
\item The support changes slowly over time, i.e. $S_a \ll |N_t|$ and $S_r \ll |N_t|$. This is empirically verified in \cite{just_lscs,kfcsmri}.%
\item The nonzero values also change slowly, i.e. $\sigma_{sys}^2$ is small.
\een
\end{ass}

\Subsection{KF CS-residual (KF-CS) algorithm}
\label{kfcsalgo}
Recall that $T :=\Nhat_{t-1}$ denotes the support estimate from $t-1$. KF CS-residual (KF-CS) runs a KF for the system in (\ref{obsmod}), (\ref{sysmod}) but with $Q_t$ replaced by $\hat{Q}_t = \sigma_{sys}^2 I_{T}$ and computes the KF residual, denoted $\tty$. The new additions, if any, to $T$, are detected by performing CS on $\tty$ and thresholding the output. If the support set changes, an LS estimate is computed using the new support estimate. If it does not change, we just use the initial KF output as the estimate. We then use this estimate to compute deletions from the support by thresholding with a different (typically larger) threshold. Once again, if the support set changes, a final LS estimate is computed using the new support and if not, then we just use the initial KF output.

In this work, the CS-residual step in KF-CS uses the Dantzig selector \cite{dantzig} (but this can be easily changed to BPDN or Lasso or any greedy method such as OMP etc), i.e. it solves
\bea
\min_{\zeta} \|\zeta\|_1 \ \text{s.t.} \ \|A'(y - A \zeta)\|_\infty < \lambda 
\label{simplecs}
\eea
with $y$ replaced by the current KF residual, $\tty$.

Let $P_{t|t-1}$, $P_t$ and $K_t$ denote the ``assumed" prediction and update error covariance matrices and the Kalman gain used by the KF in KF-CS. We say ``assumed" since the KF does not always use the correct value of $Q_t$ and so $P_{t|t-1}$ or $P_t$ are also not equal to the actual error covariances.

{\em We summarize the complete KF-CS algorithm below.}

\noindent {\em Initialization ($t=0$): } At $t=0$, we run simple CS (Dantzig selector) with a large enough number of measurements, $n_0 > n$, i.e. we solve (\ref{simplecs}) with $y=y_0$ and $A = A_0$ ($A_0$ will be an $n_0 \times m$ matrix). This is followed by support estimation and then LS estimation as in the Gauss-Dantzig selector. We denote the final output by $\xhat_0$ and its estimated support by $\Nhat_0$.
For $t>0$ do,
\ben
\item {\em Initial KF. } Let $T = \Nhat_{t-1}$. Run Kalman prediction and update using $\hat{Q}_t = \sigma_{sys}^2 I_{T}$ and compute the KF residual, $\tty$, using
\label{initkf}
\bea
P_{t|t-1} \se P_{t-1} + \hat{Q}_t, \ \text{where} \ \hat{Q}_t :=\sigma_{sys}^2 I_{T}  \nn \\ 
K_{t} \se P_{t|t-1} A' (AP_{t|t-1} A' + \sigma^2 I)^{-1}  \nn \\
P_t \se (I - K_{t} A) P_{t|t-1} \nn \\
\xhat_{t,\text{init}} \se (I - K_{t} A) \xhat_{t-1} + K_{t}  y_t \nn \\
\tty \se y_t - A \xhat_{t,\text{init}}    
\label{kftmp}
\eea

\item {\em CS-residual. } Do CS (Dantzig selector) on the KF residual, i.e.  solve (\ref{simplecs})
with $y= \tty$. Denote its output by $\betahat_t$. Compute 
\bea
\xhat_{t,\CSres} = \xhat_{t,\text{init}} + \betahat_t
\eea

\item {\em Detection and LS. } Detect additions to $T$ using
\label{det_ls_step}
\bea
\tT_\dett \se T \cup \{ i \in T^c : |(\xhat_{t,\CSres})_i| > \alpha \} \nn
\label{kfcs_detect}
\eea
If $\tT_\dett$ is equal to $T$,
 set $\xhat_{t,\dett}=\xhat_{t,\text{init}}$,
\\ else,
\\ compute an LS estimate using $\tT_\dett$, i.e. compute 
\bea
(\xhat_{t,\dett})_{\tT_\dett} \se {A_{\tT_\dett}}^\dag  y_t, \ \ (\xhat_{t,\dett})_{\tT_\dett^c} = 0
\label{det_ls}
\eea

\item {\em Deletion and Final LS. } Estimate deletions to $\tT_\dett$ using
\label{final_ls_step}
\bea
\Nhat_t =  \tT_\dett \setminus \{i \in \tT_\dett: |(\xhat_{t,\dett})_i| < \alpha_{del} \}
\eea
If $\Nhat_t$ is equal to $T$,
 set $\xhat_{t}=\xhat_{t,\text{init}}$,
\\ else,
\\ compute an LS estimate using $\Nhat_t$ and update $P_t$, i.e.
\bea
(\xhat_{t})_{\Nhat_t} \se {A_{\Nhat_t}}^\dag  y_t, \nn \\
(\xhat_{t})_{\Nhat_t^c} \se 0 \nn \\
(P_t)_{\Nhat_t,\Nhat_t} \se ({A_{\Nhat_t}}'{A_{\Nhat_t}})^{-1} \sigma^2, \nn \\
(P_t)_{\Nhat_t^c,[1,m]} \se 0, \   (P_t)_{[1,m],\Nhat_t^c} = 0
\label{final_ls}
\eea

\item Output $\xhat_t$ and $\hat{z}_t = \Phi \xhat_t$. Feedback $\xhat_t$, $P_t$, $\Nhat_t$.

\een

\begin{remark}
Notice that the final LS step re-initializes the KF whenever the estimated support changes. This ensures less dependence of the current error on the past, and makes the stability analysis easier.%
\end{remark}

\begin{remark} 
For ease of notation, in (\ref{kftmp}), we write the KF equations for the entire $x_t$. But the algorithm actually runs a reduced order KF for only $(x_t)_T$ at time $t$, i.e. we actually have $(\xhat_t)_{T^c} = 0$, $(K_t)_{T^c,[1:n]} = 0$, $(P_{t|t-1})_{[1,m],T^c} = 0$, $ (P_{t-1})_{[1,m],T^c} = 0$, $(P_{t|t-1})_{T^c,[1,m]} = 0$, and $(P_{t-1})_{T^c,[1,m]} = 0$. For computational speedup, the reduced order KF should be explicitly implemented.
\end{remark}

\begin{remark}
The KF in KF-CS does not always run with correct model parameters. Thus, even when $\sigma_{sys}^2/\sigma^2$ is small, it is not clear if KF-CS will always outperform LS-CS \cite{just_lscs}. This will hold at times when the support is accurately estimated and the KF has stabilized [see Fig. \ref{kfcs_better_59}]. Also, this will hold when support changes occur slowly enough, and $n$ is small so that LS-CS error becomes instable, but is just large enough to prevent KF-CS instability [see Fig. \ref{kfcs_better_40}].
\end{remark}

\Subsection{Discussion of the Signal Model}
\label{whyrandomwalk}
A more accurate model than Signal Model \ref{randomwalk} would be random walk with nonzero and time-varying drift. If accurate knowledge of the time-varying drift is available, the KF estimation error can be reduced significantly. But, in practice, to estimate the time-varying drift values, one would need a large number of identically distributed training signal sequences, which is an impractical assumption in most cases. On the other hand, in the above model the parameters are time-invariant and their values can be estimated from a single training sequence. This is done in \cite{kfcsmri,chenlu_tip}.

Now, a random walk model at all times is not a realistic signal model since it implies that the signal power keeps increasing over time. The following is what is more realistic. A new sparse basis coefficient starts from zero and slowly increases to a certain roughly constant value, i.e. it follows a random walk model for sometime and then reaches steady state. Steady state can usually be accurately modeled by a (statistically) stationary model with nonzero mean. To design KF-CS for such a model one would either need to detect when a coefficient becomes stationary or one would need to know it ahead of time. The former will typically be very error prone while the latter is an impractical assumption. To avoid having to do this, we just assume a random walk model at all times.

In Sec. \ref{sims}, we show that the KF-CS algorithm of Sec. \ref{kfcsalgo} works both for data generated from  Signal Model \ref{randomwalk} and for data generated from a more realistic bounded signal power model taken from \cite{just_lscs}, which is a deterministic version of what is discussed above. In \cite{kfcsmri,chenlu_tip}, we show that it works even for actual image sequences.


\Section{Analyzing (KF)CS-residual step}
\label{whycsresworks}
The KF residual, $\tty$, can be rewritten as $\tty = A \beta_t + w_t$ where
\bea
(\beta_t)_\Delta \se (x_t - \xhat_{t,\text{init}})_\Delta = (x_t)_\Delta \nn \\
(\beta_t)_T \se (x_t - \xhat_{t,\text{init}})_T \nn \\
          \se [I - K_t A_T](x_{t} - \xhat_{t-1})_T - K_t A_\Delta (x_t)_\Delta - K_t w_t \nn \\
(\beta_t)_{(T \cup \Delta)^c} \se 0
\label{defbeta}
\eea
where $T = \Nhat_{t-1}$ and $K_t \equiv  (K_t)_{T,[1,n]}$. Thus, $\beta_t$ is $|T \cup \Delta|=|N_t \cup \Delta_e|$ sparse.

In Appendix \ref{kfcsresbnd}, we show that $\|(\beta_t)_T\|$ is bounded as in (\ref{betabound}).
As we argue there, if (a) the support changes slowly enough, (b) the signal values change slowly enough, (c) the noise is small enough and (d) the previous reconstruction is accurate enough, this bound will be small, i.e. $\beta_t$ will be compressible along $T$. In other words, $\beta_t$ will be only $|\Delta|$-approximately-sparse. Because of (a) and (d), $|\Delta|$ will be small compared to $|N_t|$. Thus doing CS on $\tty$ will incur much less error than doing CS on $y_t$ (simple CS), which needs to reconstruct a $|N_t|$-sparse signal, $x_t$. This statement can be quantified by using (\ref{betabound}) to bound CS-residual error exactly like in \cite[Theorem 1]{just_lscs} and then doing the comparison with CS also as in \cite{just_lscs}.

The CS-residual error bound will be directly proportional to the bound on $\|(\beta_t)_T\|$ given in (\ref{betabound}). This can be used to argue why KF-CS outperforms LS-CS when $n$ is smaller and support changes slowly enough. We do this in Appendix \ref{kfcs_lscs}.

\Section{KF-CS Error Stability}
\label{supplem_kfcsstab}
Analyzing the KF-CS algorithm of Sec. \ref{kfcsalgo}, which includes the deletion step, is difficult using the approach that we outline below. Thus, in this section, we study KF-CS without the deletion step, i.e. we set $\alpha_{del}=0$. KF-CS without deletion assumes that there are few and bounded number of removals and false detects. For simplicity, in this work, we just assume $S_r=0$ in Signal Model \ref{randomwalk} and we will select $\alpha$ so that there are zero false detects. $S_r=0$ along with the assumption that the maximum sparsity size is $\smax$ implies that there are only a finite number of addition times, $K$, i.e. for all  $t \ge t_{K-1}$, $N_t = N_{t_{K-1}}$. We summarize this in the following signal model.
\begin{sigmodel}
Assume Signal Model \ref{randomwalk} with $S_r=0$. This implies that there are only a finite number of addition times, $t_j$, $j=0, 1, \dots (K-1)$ and $K = \lceil \frac{\smax - S_0}{S_a} \rceil$. Let $t_K:=\infty$.
\label{randomwalk_norems}
\end{sigmodel}

Consider the genie-aided KF, i.e. a KF which knows the true support $N_t$ at each $t$. It is the  MMSE estimator of $x_t$ from $y_1, \dots y_t$ if the support sets, $N_t$, are assumed known and the noise is Gaussian, and is the linear MMSE for any arbitrary noise. In this section, we find sufficient conditions under which, with high probability (w.h.p.), KF-CS for Signal Model \ref{randomwalk_norems} and observation model given by (\ref{obsmod}) gets to within a small error of the genie-KF for the same system, within a finite delay of the new addition time. Since the genie-KF error is itself stable w.h.p., as long as $\delta_{\smax}<1$, this also means that the KF-CS reconstruction error is stable w.h.p.%

Our approach involves two steps. Consider $t \in [t_j,t_{j+1})$. First, we find the conditions under which w.h.p. all elements of the current support, $N_t = N_{t_j}$ get detected before the next addition time, $t_{j+1}$. Denote the detection delay by $\tau_\dett$. If this happens, then during $[t_j+\tau_\dett,t_{j+1})$, both KF-CS and genie-KF run the same fixed dimensional and fixed parameter KF, but with different initial conditions. Next, we show that if this interval is large enough, then, w.h.p, KF-CS will stabilize to within a small error of the genie-KF within a finite delay after $t_j+\tau_\dett$. Combining these two results gives our stability result.

We are able to do the second step because, whenever $\Nhat_t \neq \Nhat_{t-1}$, the final LS step re-initializes the KF with $P_t$, $\xhat_t$ given by (\ref{final_ls}). This ensures that the KF-CS estimate, $\xhat_t$, and the Kalman gain, $K_t$, at $t+1$ and future times depend on the past observations only through $T:=\Nhat_{t}$. Thus, conditioned on the event $\{\Nhat_t = N_t, \ \forall  \ t \in [t_j+\tau_\dett,t_{j+1}) \}$, there will be no dependence of either $\xhat_t$ or of $K_t$ on observations before $t_j+\tau_\dett$.

\Subsection{The Stability Result}
We begin by stating Lemma \ref{finitedelay} which shows two things. First, if accurate initialization is assumed, the noise is bounded, $\smax \le \sinf$,  $\alpha_{del}=0$  and $\alpha$ is high enough, there are no false detections. If the delay between addition times also satisfies $d > \tau_{\dett}(\eps,S_a)$, where $ \tau_{\dett}$ is what we call the ``high probability detection delay", then the following holds. If before $t_j$, the support was perfectly estimated, then w.p. $\ge 1-\eps$, all the additions which occurred at $t_j$ will get detected by $t_j + \tau_{\dett}(\eps,S_a) < t_{j+1}$.


\begin{lemma}
Assume that $x_t$ follows Signal Model \ref{randomwalk_norems}. If
\ben
\item {\em (initialization ($t=0$))} all elements of $x_0$ get correctly detected and there are no false detects, i.e. $\Nhat_0 = N_0$,
\label{initass}

\item {\em (measurements)} $\smax \le \sinf$ and $\|w\|_\infty \le \lambda/\|A\|_1$,
\label{measmod}

\item {\em (algorithm)} we set $\alpha_{del}=0$ and $\alpha^2  = B_* := C_1(\smax) \smax \lambda^2$, where $C_1(S)$ is defined in \cite[Theorem 1.1]{dantzig},%
\label{thresholds}

\item  {\em (signal model)} delay between addition times, $d > \tau_{\dett}(\eps,S_a)$, 
\bea
\text{where~} \tau_{\dett}(\eps,S):= \left\lceil  \frac{4 B_*}{ \sigma_{sys}^2 [\Q^{-1}(\frac{ (1- \eps)^{1/S} }{2})]^2 } \right\rceil - 1,
\label{defteps}
\eea
 $\lceil \cdot \rceil$ denotes the greatest integer function and $\Q(z):= \int_{z}^\infty (1/\sqrt{2\pi}) e^{-x^2/2} dx$ is the Gaussian Q-function,%
\een
then
\ben
\item at each $t$, $\Nhat_t \subseteq N_t \subseteq N_{t+1}$ and so $|\Delta_{e,t+1}|=0$
\item at each $t$, $\| x_t - \xhat_{t,\CSres}\|^2 \le B_*$
\item $Pr(E_j| F_{j}) \ge 1-\eps$ where $F_{j}:= \{\Nhat_t = N_t \ \text{for} \ t=t_{j}-1 \}$ and $E_j:= \{\Nhat_t = N_t, \ \forall \ t \in [t_j + \tau_{\dett}(\eps,S), t_{j+1}-1] \}$.
\een
\label{finitedelay}
\end{lemma}

The proof is given in Appendix \ref{finitedelay_proof}. The initialization assumption is made only for simplicity. It can be easily satisfied by using $n_0 > n$ to be large enough. Next we give Lemma \ref{kfinitwrong} which states that if the true support set does not change after a certain time, $t_{nc}$, and if it gets correctly detected by a certain time, $t_* \ge t_{nc}$, then KF-CS converges to the genie-KF in mean-square and hence also in probability.
\begin{lemma}
Assume that $x_t$ follows Signal Model \ref{randomwalk_norems}; $\delta_{\smax} < 1$; and $\alpha_{del}=0$. 
Define the event $D:=\{\Nhat_t = N_t = N_*, \ \forall \ t \ge t_* \}$. Conditioned on $D$, the difference between the KF-CS estimate, $\xhat_t$ and the genie-aided KF estimate, $\xhat_{t,GAKF}$, $\diff_t :=\xhat_{t}-\xhat_{t,GAKF}$, converges to zero in mean square and hence also in probability. $\blacksquare$
\label{kfinitwrong}
\end{lemma}

The proof is similar to what we think should be a standard result for a KF with wrong initial conditions (here, KF-CS with $t=t_*$ as the initial time) to converge to a KF with correct initial conditions (here, genie-KF) in mean square.  A similar (actually stronger) result is proved for the continuous time KF in \cite{ocone}. We could not find an appropriate citation for the discrete time KF and hence we just give our proof in Appendix \ref{kfinitwrongproof}. After review, this can be significantly shortened. The proof involves two parts. First, we use the results from \cite{ekf} and \cite{hornjohnson} to show that (a) $P_{t|t-1}, P_t, K_t$ and $J_t:= I - K_t A_{N_*}$ converge to steady state values which are the same as those for the corresponding genie-KF; and (b) the steady state value of $J_t$, denoted $J_*$, has spectral radius less than 1 and because of this, there exists a matrix norm, denoted $\|.\|_\rho$, s.t. $\|J_*\|_\rho < 1$. Second, we use (a) and (b) to show that the difference in the KF-CS and genie-KF estimates, $\diff_t$, converges to zero in mean square, and hence also in probability (by Markov's inequality).


A direct corollary of the above lemma is the following.
\begin{corollary}
Assume that $x_t$ follows Signal Model \ref{randomwalk_norems}; $\delta_{\smax} < 1$; and $\alpha_{del}=0$. 
Define the event $D_f:=\{\Nhat_t = N_t = N_*, \ \forall \ t \in [t_*, \ t_{**}] \}$.
For a given $\eps, \eps_{\text{err}}$, there exists a $\tau_{KF}(\eps,\eps_{\text{err}},N_*)$ s.t. for all $t \in [t_* + \tau_{KF} , \ t_{**}]$, $Pr(\|\diff_t\|^2 \le \eps_{\text{err}} \ | \ D_f) > 1- \eps$. Clearly if $t_{**} < t_* + \tau_{KF}$, this is an empty interval.
\label{kfinitwrong_cor}
\end{corollary}

The stability result then follows by applying Lemma \ref{kfinitwrong} followed by Corollary \ref{kfinitwrong_cor} for each addition time, $t_j$.%
\begin{theorem}[KF-CS Stability]
Assume that $x_t$ follows Signal Model \ref{randomwalk_norems}.
Let  $\diff_t:=\xhat_{t}-\xhat_{t,GAKF}$ where $\xhat_{t,GAKF}$ is the genie-aided KF estimate and $\xhat_t$ is the KF-CS estimate. For a given $\eps, \eps_{\text{err}}$, if the conditions of Lemma \ref{finitedelay} hold, and if the delay between addition times, $d > \tau_{\dett}(\eps,S_a)+ \tau_{KF}(\eps,\eps_{\text{err}},N_{t_j})$, where $\tau_{\dett}(.,.)$ is defined in (\ref{defteps}) in Lemma \ref{finitedelay} and $\tau_{KF}(.,.,.)$ in Corollary \ref{kfinitwrong_cor},
then
\ben
\item  $Pr(\|\diff_t\|^2 \le \eps_{\text{err}}) > (1-\eps)^{j+2}$, for all $t \in [t_j + \tau_{\dett}(\eps,S_a) + \tau_{KF}(\eps,\eps_{\text{err}},N_{t_j}), \ t_{j+1}-1]$, for all $j=0,\dots (K-1)$,
\item $Pr(|\Delta| \le S_a \text{~and~} |\Delta_{e}|=0, \ \forall \ t) \ge (1-\eps)^K$
\item $Pr(|\Delta|=0 \text{~and~} |\Delta_{e}|=0, \ \forall \ t \in [t_j + \tau_{\dett}(\eps,S_a), t_{j+1}-1], \ \forall \ j=0,\dots K-1) \ge (1-\eps)^K$.

\een
\label{kfcs_stab}
\end{theorem}
The proof is given in Appendix \ref{append_kfcs}.
A direct corollary is that after $t_{K-1}$ KF-CS will converge to the genie-KF in probability. This is because for $t \ge t_{K-1}$, $N_t$ remains constant ($t_{K}=\infty$).



\subsection{Discussion}
Consider a $t \in [t_j,t_{j+1})$. Notice that $\tau_{KF}$ depends on the current support, $N_t=N_{t_j}$ while $\tau_\dett$ depends only on the number of additions at $t_j$, $S_a$. Theorem \ref{kfcs_stab} says that if $n$ is large enough so that $\smax \le \sinf$; $\alpha_{del}=0$ (ensures no deletions); $\alpha = \sqrt{B_*}$  (ensures no false detects); and if the time needed for the current KF to stabilize, $\tau_{KF}(\eps,\eps_{\text{err}},N_{t_j})$, plus the high probability detection delay,  $\tau_{\dett}(\eps,S_a)$, is smaller than $d$, then w.p. $\ge (1-\eps)^{j+2}$, KF-CS will stabilize to within a small error, $\eps_{\text{err}}$, of the genie-KF before the next addition time, $t_{j+1}$. If the current $\tau_{KF}$ is too large, this cannot be claimed.
But as long as $\tau_{\dett}(\eps,S_a) < d$, the unknown support size, $|\Delta|$ remains bounded by $S_a$, w.p. $\ge (1-\eps)^K$.

We give our result for the case of zero removals and zero false detects, but the same idea will extend even if $|\Delta_e|$ is just bounded.

As explained in Sec. \ref{whyrandomwalk}, most signals do not follow a random walk model forever (such a model would imply unbounded signal power).
In practice, a new coefficient may start with following a random walk model, but eventually reach steady state (stationary model). In this case, it should be possible to modify our result to claim that if, before reaching steady state, all coefficients become large enough to exceed the threshold plus upper bound on error, and if this happens before the next addition time, KF-CS remains stable.

Our result is weaker than that of LS-CS \cite{just_lscs} - it needs $\smax \le \sinf$ (the LS-CS result only needs $S_a \le \sinf$ and $\smax \le \sone$); it uses a random walk model; it does not handle support removals; and the computed high-probability detection delay is quite loose\footnote{Our result may even go through if CS-residual was replaced by CS.}. This is due to two main reasons. One is that we assume a zero drift random walk model as the signal model both for defining KF-CS and for analyzing it, while LS-CS uses a model with nonzero drift for the analysis (the algorithm does not assume any signal model). The reason for our choosing this model is explained in Sec. \ref{whyrandomwalk}. The second and more important reason is that bounding KF error is more difficult than bounding LS error. This is because the KF error, and hence also the (KF)CS-residual error, depends on the previous reconstruction error. The (LS)CS-residual error only depends on $|T|$, $|\Delta|$ and if we can get a time-invariant bound on these, we can do the same for the error.

\begin{figure}[t!]
\centerline{
\subfigure[Signal Model \ref{randomwalk_norems} with $S_0=8$, $S_a=2$, $d=5$, $\smax=26$]{
\label{oldsim}
\epsfig{file = 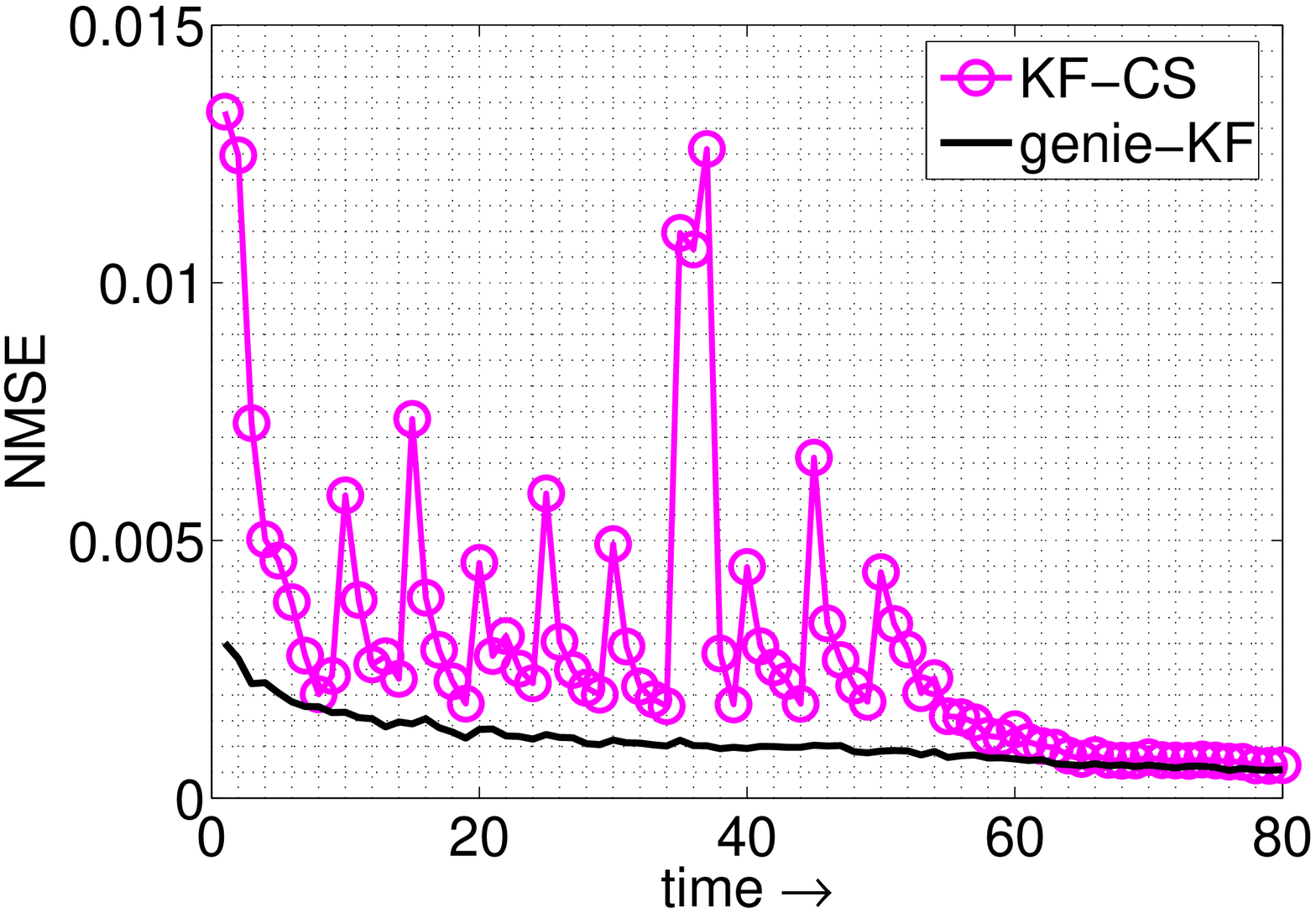, height = 3cm,width=4cm}
}
\subfigure[Signal Model \ref{randomwalk_norems} with $S_0=8$, $S_a=4$, $d=10$, $\smax=20$]{
\label{oldsim2}
\epsfig{file =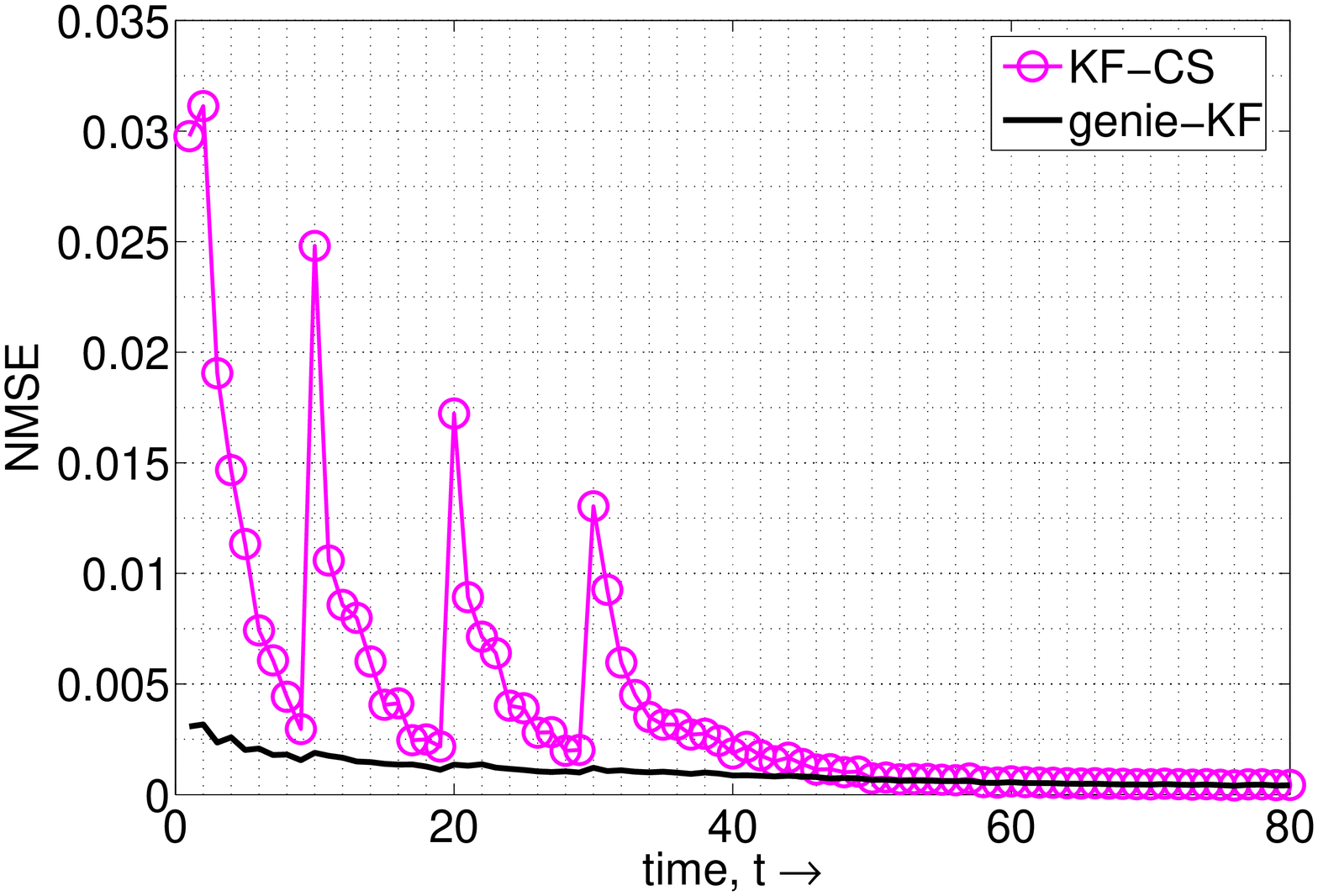, height = 3cm,width=4cm}
}
}
\vspace{-0.05in}
\caption{\small{Verifying KF-CS stability for Signal Model \ref{randomwalk_norems}.}}
\vspace{-0.2in}
\label{kfcs_stab}
\end{figure}

\begin{figure*}[t!]
\centerline{
\subfigure[$n = 59$, NMSE]{
\label{kfcs_better_59}
\epsfig{file = 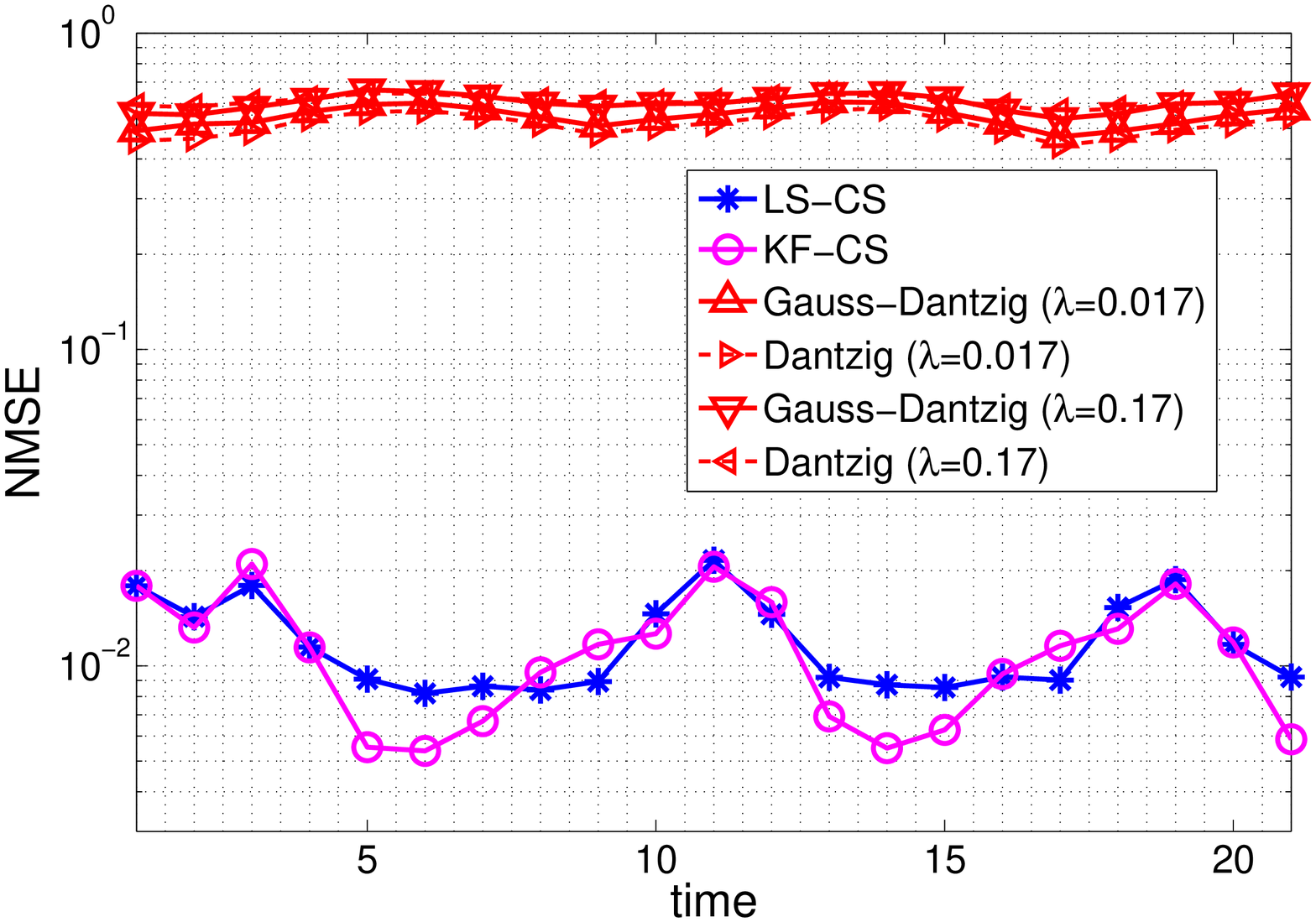, height = 3cm,width=4cm}
}
\subfigure[$n = 59$, misses \& extras]{
\label{kfcs_better_59_support}
\epsfig{file = 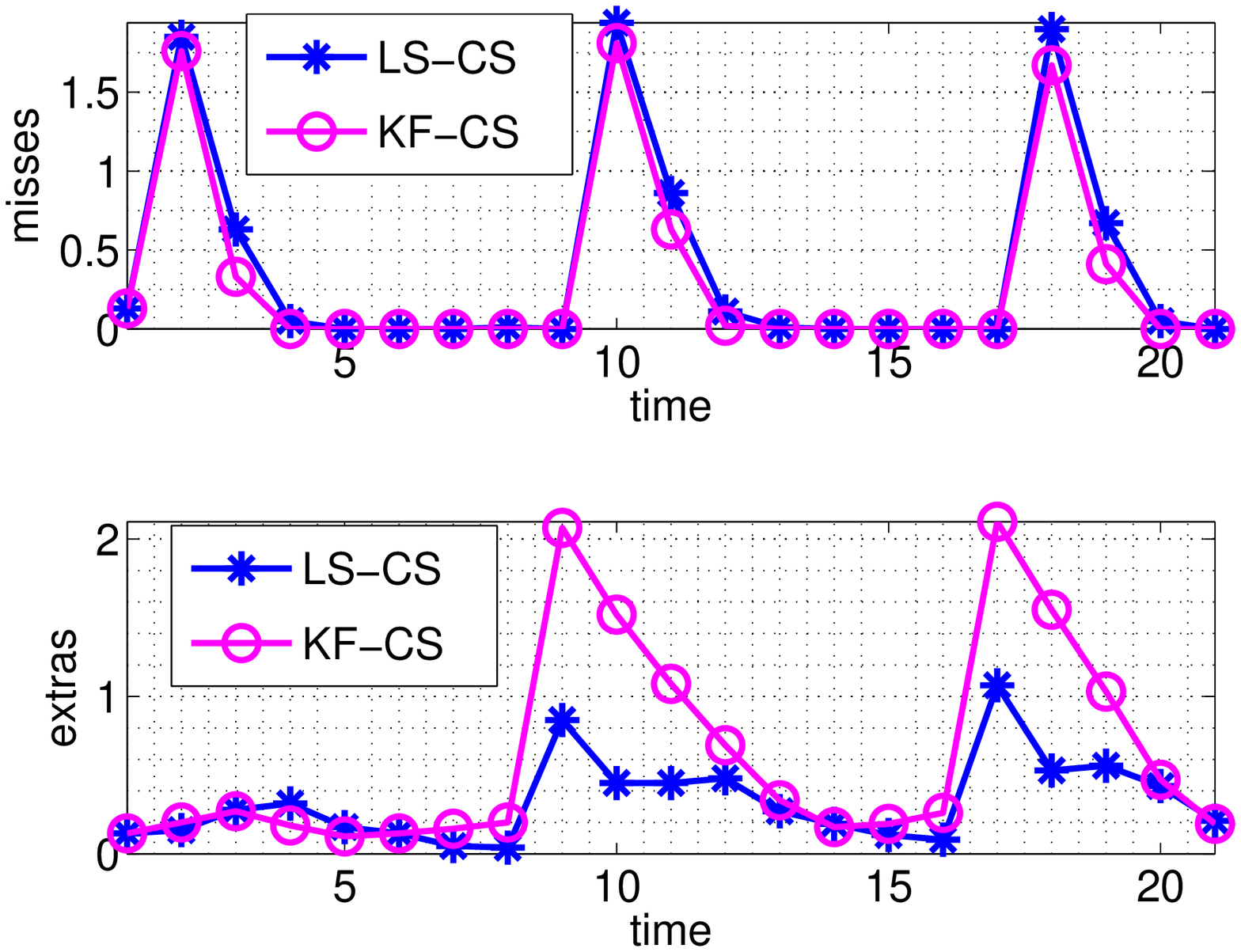, height = 3cm,width=4cm}
}
\subfigure[$n = 45$, NMSE]{
\label{kfcs_better_40}
\epsfig{file = 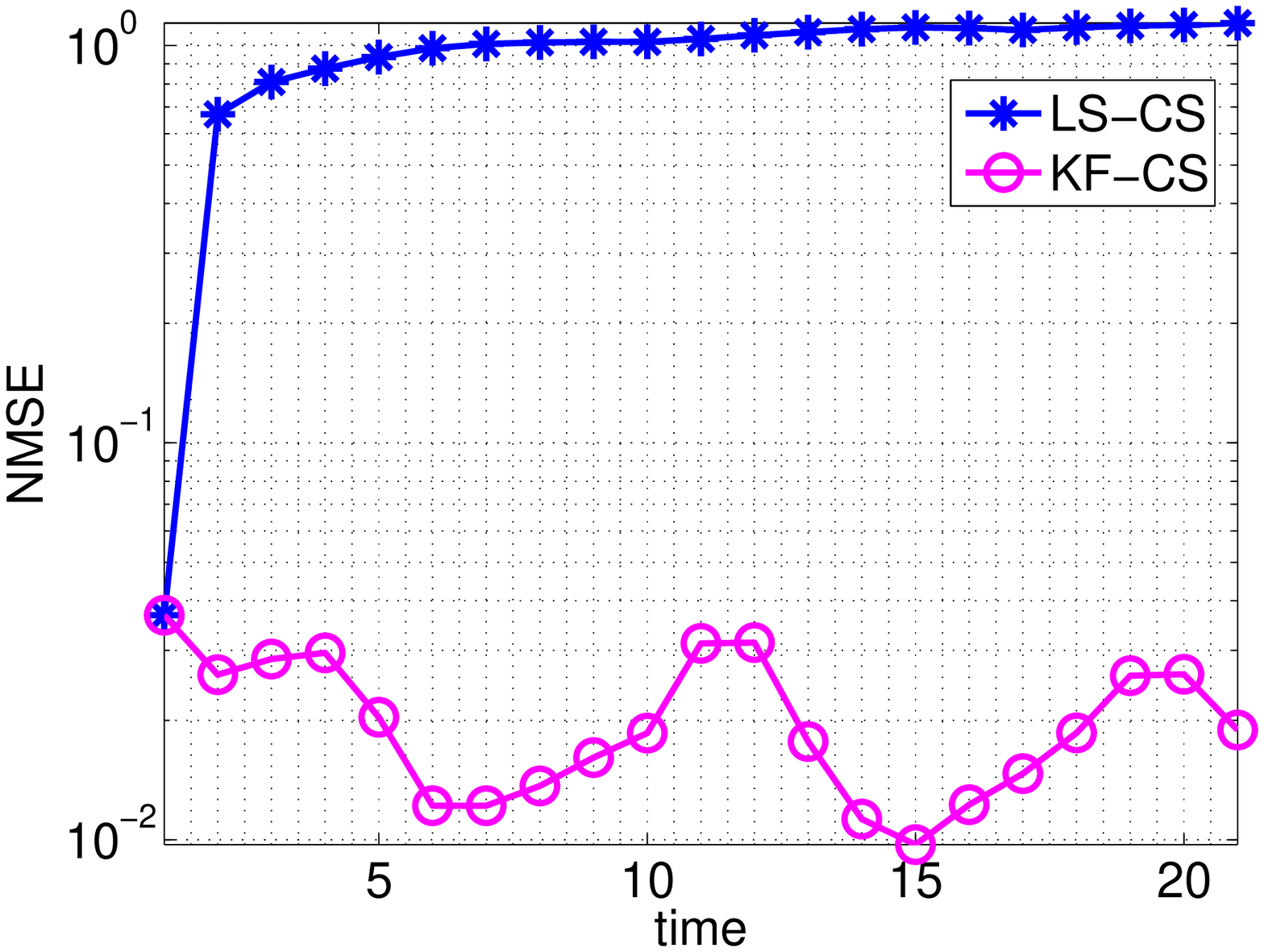, height = 3cm,width=4cm}
}
\subfigure[$n=45$, misses \& extras]{
\label{kfcs_better_40_support}
\epsfig{file = 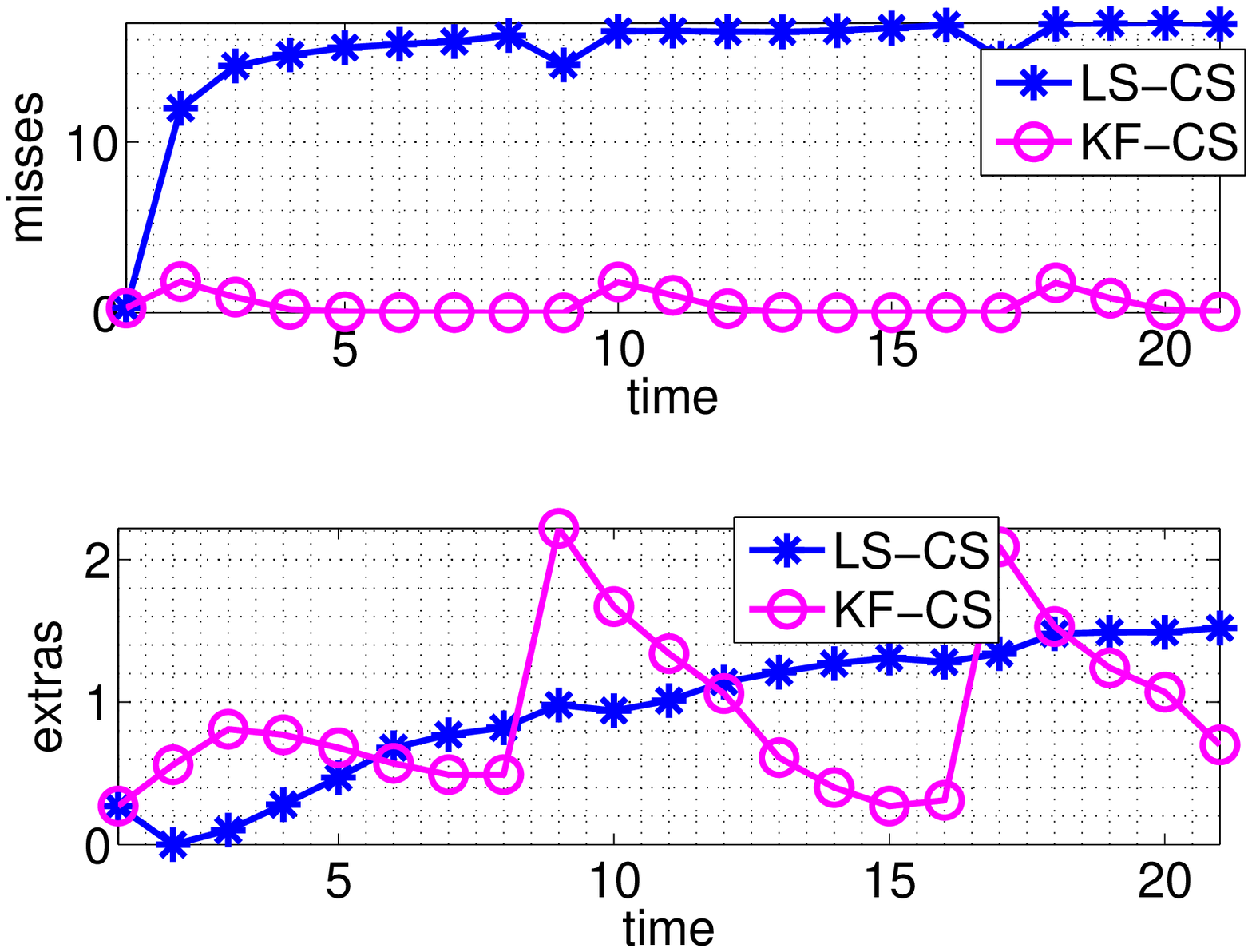, height = 3cm,width=4cm}
}
}
\vspace{-0.05in}
\caption{\small{Comparing KF-CS with CS and LS-CS. CS-residual in LS-CS or in KF-CS used $\lambda=0.17$. Misses = $\E[|N_t \setminus \Nhat_t|]$, Extras = $\E[|\Nhat_t \setminus N_t|]$.%
%
}}
\vspace{-0.2in}
\label{lscs_kfcs}
\end{figure*}






\Section{Simulation Results}
\label{sims}
We discuss two sets of simulation results. The first simulates data according to Signal Model \ref{randomwalk_norems} and verifies KF-CS stability. The second set of simulations compares KF-CS with LS-CS \cite{just_lscs} and simple CS (Dantzig selector) \cite{dantzig}. This comparison uses the more realistic signal model assumed in \cite{just_lscs}, which has a roughly constant signal power and support size and allows regular additions and removals from support. 

\Subsection{Signal Model \ref{randomwalk_norems}: verify KF-CS Stability}

We simulated Signal Model \ref{randomwalk_norems} with $m=256$, $S_0=8$, $S_a=2$, $d=5$, $\smax = 26$ and $\sigma_{sys}=1$. Thus additions occurred at $t=1,6,11,\dots,46$. The measurement model used $n=n_0=72$ and Gaussian noise with $\sigma=0.16$. The normalized MSE (NMSE) is plotted in Fig. \ref{oldsim}. In a second simulation, we increased $S_a$, but we also increased $d$: we used $S_a=4$, $d=10$ and $\smax = 20$ and everything else was the same.
We show the error plot in Fig. \ref{oldsim2}. Notice that in both cases, (i) KF-CS stabilizes to within a small error of the genie-KF within a short delay of a new addition time; and (ii) after the final set of new additions, KF-CS converges to the genie-KF. The difference between the two is that the peak errors at the new addition time are larger in the second case (since $S_a$ is larger).%

We implemented the KF-CS algorithm of Sec. \ref{kfcsalgo} but without the deletion step, i.e. we set $\alpha_{del}=0$. Since the observation noise is not truncated, occasionally the addition step can result in a very large number of false additions. To prevent this, we restricted the maximum number of allowed additions at a given time to $\gamma n/\log_2 m$ ($\gamma$ between 0.7 and 1.25) largest magnitude coefficients.


\Subsection{Bounded signal power model from \cite{just_lscs}}
\label{lowsnrsims}
For this comparison we used the signal model of \cite{just_lscs}. This is a realistic signal model with roughly constant signal power and support size. We used $m=200$, $S_0=20$, $S_a=2=S_r$, $a_i = 0.2$, $M=1$, $d=8$ and $r=3$. Thus new additions occurred at $t=2,10,18$. Coefficient decrease began at $t=7,15$ and these got removed at $t=9,17$ respectively. The measurement noise was $uniform(-c,c)$.

In the first simulation, we used $n_0=150$, $n=59$ and $c=0.1266$. LS-CS used $\lambda = 0.176$, $\alpha=c/2 =0.06= \alpha_{del}$. Also, it restricted maximum number of additions at a time to $S_a+1$. The KF-CS algorithm of Sec. \ref{kfcsalgo} was implemented. It used the above parameters and it set $\sigma^2=c^2$ and $\sigma_{sys}^2 = 0.01$. For the signal model of \cite{just_lscs}, there are no correct choices of KF parameters. The average of $(x_t-x_{t-1})_i^2$ over $i$ and $t$ was $(0.04*(5/8)*(2/20) + 0.11*(3/8)*(2/20) +0*1*(16/20)) \approx 0.01$ and this motivated the choice of  $\sigma_{sys}^2$. The noise variance is $c^2/3$, but we use a larger value to also model the effect of extra observation error due to the unknown support $\Delta$. 
%
%
The NMSE plot is shown in Fig. \ref{kfcs_better_59}. The mean number of misses ($\E[|N_t \setminus \Nhat_t|]$) and of extras ($\E[|\Nhat_t \setminus N_t|]$) are plotted in Fig. \ref{kfcs_better_59_support}. We averaged over 100 Monte Carlo runs. Notice that right after a new addition, both LS-CS and KF-CS have similar MSE, but in the stable state KF-CS stabilizes to a smaller value. The NMSEs for CS (Dantzig selector) and Gauss-Dantzig selector even with different choices of $\lambda$ are much larger (40-60\%).

In a second simulation, we used $n_0=150$, $n=45$ and $c=0.15$ and everything else was the same as above. The error plots are shown in Fig. \ref{kfcs_better_40} and the number of extras and misses are plotted in Fig. \ref{kfcs_better_40_support}. With such a small $n$, LS-CS error becomes instable. But $n=45$ (along with large enough delay between addition times, $d=8$ and small enough $r=\sigma_{sys}^2/\sigma^2 = 0.44$) is large enough to prevent KF-CS instability.%



\Section{Conclusions and Future Work}
\label{conclusions}

We proposed KF CS-residual (KF-CS) which replaces CS on the raw observation by CS on the KF residual, computed using the known part of the support. We proved KF-CS stability, but the assumptions used were somewhat strong (stronger than those used for LS-CS \cite{just_lscs}). We demonstrated via simulations that KF-CS error is stable and small under much weaker assumptions. Also, it significantly outperformed LS-CS when the available number of measurements was very small.

A key direction of future work is to prove KF-CS stability under weaker assumptions. This will require assuming a signal model with nonzero drift (to get a tighter detection delay bound) and bounded signal power. It may also help to assume a statistical prior on support change, e.g. by using a model similar to \cite{schniter}.
A useful extension of KF-CS would be to replace CS-residual by modified-CS \cite{isitmodcs}.

\appendix

\Subsection{Bounding $\|(\beta_t)_T\|$}
\label{kfcsresbnd}
Recall that $T_t =\Nhat_{t-1}$ and $\Delta_t = N_t \setminus \Nhat_{t-1}$. Let $\delta_t \defn \delta_{|T_t|}$ and $\theta_t \defn \theta_{|T_t|,|\Delta_t|}$. Also, let  $K_t \equiv (K_t)_{T,[1,n]}$,
\bea
M_t \sdefn {A_T}'A_T + (P_{t|t-1})_{T,T}^{-1} \sigma^2 \text{~and~} \nn \\
r \sdefn \sigma_{sys}^2/\sigma^2
\label{defr}
\eea
We use the following simple facts in the discussion below \cite{hornjohnson}. For symmetric positive definite matrices, $M$, $\tilde{M}$, $\|M\| = \lambda_{\max}(M) = 1/\lambda_{\min}(M^{-1})$,  $\lambda_{\min}(M+\tilde{M}) \ge  \lambda_{\min}(M) + \lambda_{\min}(\tilde{M})$ while the inequality holds in the opposite direction for  $\lambda_{\max}$. Here $\lambda_{\max}(M)$, $\lambda_{\min}(M)$ denote the maximum, minimum eigenvalue of $M$.

As is well known \cite{ekf}, $K_t$, $[I - K_t A_T]$, $P_t$ can be rewritten as
\bea
K_t \se M_t^{-1} {A_T}' \nn \\
J_t := I - K_t A_T \se M_t^{-1} (P_{t|t-1})_{T,T}^{-1} \sigma^2 \nn \\
(P_{t|t-1})_{T,T} \se (P_{t-1})_{T,T} + (\sigma_{sys}^2 I_{T})_{T,T}, \ \text{where} \nn \\
(P_{t-1})_{T,T} \se \left\{ \begin{array}{cc}
                  M_{t-1}^{-1} \sigma^2 & \text{~if~} T_{t}=T_{t-1} \\
                  ({A_{T_{t}}}'{A_{T_{t}}})^{-1} \sigma^2 & \text{~if~} T_{t} \neq T_{t-1}
                  \end{array} \right.
\label{Ktrewrite}
\eea
The third equation is repeated from (\ref{kftmp}). 
To bound $\|(\beta_t)_{T}\|$, defined in (\ref{defbeta}), we need to bound $\|J_t\|$ and $\|K_t {A_T}'A_\Delta\|$, which in turn requires bounding $\|M_t^{-1}\|$, $\|(P_{t|t-1})_{T,T}^{-1}\|$ and $\|{A_T}'A_\Delta\|$. Using the definition of $\theta_{S,S'}$ \cite[eq 1.5]{dantzig}, it is easy to see that $\|{A_T}'A_\Delta\| \le \theta_t$. Using (\ref{Ktrewrite}), $\|(P_{t|t-1})_{T,T}^{-1}\| \le (\lambda_{\min}(M_{t-1}^{-1})\sigma^2 + \sigma_{sys}^2)^{-1}$ if $T_{t}=T_{t-1}$ and $\|(P_{t|t-1})_{T,T}^{-1}\| \le ((1+\delta_{t})^{-1}\sigma^2 + \sigma_{sys}^2)^{-1}$ otherwise.
Also, $\|M_t^{-1}\| = \lambda_{\max}(M_t^{-1})$. Thus bounding them requires upper bounding $\lambda_{\max}(M_t^{-1})$ and lower bounding $\lambda_{\min}(M_t^{-1})$. Using the definition of the RIP constant \cite[eq. 1.3]{dantzig},
\bea
\|M_t^{-1}\| = \lambda_{\max}(M_t^{-1})
\se \frac{1}{\lambda_{\min}({A_T}'A_T + (P_{t|t-1})_{T,T}^{-1} \sigma^2)} \nn \\
 \sle \frac{1}{1-\delta_t + \frac{\sigma^2}{\lambda_{\max}((P_{t|t-1})_{T,T})}} \nn \\
 \sle \left\{  \begin{array}{cc}
  \frac{1}{1-\delta_t + \frac{1}{ \|M_{t-1}^{-1}\| + r}}  & \text{~if~} T_{t}=T_{t-1}   \\
  \frac{1}{1-\delta_t + \frac{1}{ (1-\delta_{t})^{-1} + r}}    & \text{~if~} T_{t} \neq T_{t-1}
  \end{array} \right. \nn \\
&&  \defn a_t
\label{Mtbnd}
\eea
Similarly, we can lower bound $\lambda_{\min}(M_{t-1}^{-1})$ and use it to get 
\setlength{\arraycolsep}{0.01cm}
\bea
\|(P_{t|t-1})_{T,T}^{-1}\|\sigma^2 \sle 
 \left\{  \begin{array}{cc}
\frac{1}{\frac{1}{1 + \delta_{t-1} + \frac{1}{\|M_{t-2}\|^{-1} + r}} + r} &  \text{~if~}  T_{t}=T_{t-1}=T_{t-2} \\  
\frac{1}{\frac{1}{1 + \delta_{t-1} + \frac{1}{(1+\delta_{t-1})^{-1} + r}} + r} &  \text{~if~}  T_{t}=T_{t-1} \neq T_{t-2}   \\
\frac{1}{\frac{1}{1 + \delta_{t}} + r} & \text{~if~}  T_{t} \neq T_{t-1}
\end{array} \right. \nn \\
&& \defn \frac{1}{b_t}
\label{Ppredbnd}
\eea
From (\ref{defbeta}), (\ref{Ktrewrite}), $\|(\beta_t)_T\| \le \|M_t^{-1}\| \ [ \|(P_{t|t-1})_{T,T}^{-1}\|\sigma^2 {\|(x_{t} - \xhat_{t-1})_T\|} \\ + \theta_t \|(x_t)_\Delta\| + \|{A_T}'w_t\|]$. Using this and the above bounds, we get
\bea
&& \|(\beta_t)_T\| \le a_t \left[\text{T1} + \theta_t \|(x_t)_{\Delta}\|  +  \|{A_T}'w_t\|  \right], \ \text{where} \nn \\
&& \text{T1} \defn \frac{\|(x_{t-1} - \xhat_{t-1})_{T \cap N_t} \| + \|(\xhat_{t-1})_{\Delta_e}\| + \sqrt{|T \cap N_t|} \|\nu_t\|_\infty}{b_t}, \ \ \ \ \
\label{betabound}
\eea
and $a_t$ is defined in (\ref{Mtbnd}) and $b_t$ in (\ref{Ppredbnd}).
Notice that $a_t$ is an increasing function of $\delta_t$ and $r$, and also of $\|M_{t-1}^{-1}\| \le a_{t-1}$ if $T_t=T_{t-1}$.


Now, $\Delta \subseteq (N_{t-1} \setminus T) \cup (N_t \setminus N_{t-1})$ and $\Delta_e \subseteq (T \setminus N_{t-1}) \cup (N_{t-1} \setminus N_t)$. If the previous reconstruction is accurate enough, the previous support estimate will also be accurate enough. This combined with the slow support change assumption will imply that $|\Delta|$ and $|\Delta_e|$ are small enough. $|\Delta_e|$ small enough will imply that $|T|$ is small enough (since $|T| \le |N_t| + |\Delta_e|$) and hence $\delta_{t}$ is small enough. $\delta_t$ small ensures smaller $a_t$ and larger $b_t$. $|\Delta_e|$ and $|\Delta|$ small enough will also imply that $\theta_{t}$ is small enough. The noise being small along with $|\Delta_e|$ small will imply that $\|{A_T}'w_t\|$ is small.

Slow signal value change implies (i) $r$ is small enough and (ii) at all $t$, $\|\nu_t\|_\infty$ is small enough w.h.p.. Small $r$ implies that $a_t$ is small, but it also implies that $b_t$ is small. Small $\|\nu_{t}\|_\infty$ at all $t$, along with small noise, also results in the previous reconstruction being accurate enough which, in turn means $\|(x_{t-1} - \xhat_{t-1})_{T \cap N_t} \|$ is small. Using this and the fact that only small coefficients get falsely deleted or removed\footnote{The fact that only small coefficients get removed from $N_t$ is not modeled in Signal Model \ref{randomwalk}, but is true in practice. But it is modeled in our simulations.}, $\|(\xhat_{t-1})_{\Delta_e}\|$ is also small. All this ensures that $\text{T1}$ in (\ref{betabound}) is not very large even when $b_t$ is small.
This combined with the discussion of the previous paras ensures that the bound on $\|(\beta_t)_T\|$ is small. Thus, if (a), (b), (c), (d) given in Sec. \ref{whycsresworks} hold, $\|(\beta_t)_T\|$ is small, i.e. $\beta_t$ is only $|\Delta|$-approximately-sparse; and $|\Delta|$ is small.

\Subsection{Comparing KF-CS and LS-CS using the bound on  $\|(\beta_t)_T\|$}
\label{kfcs_lscs}
 We will mention that we are only comparing upper bounds here.

Consider $a_t$ defined in (\ref{Mtbnd}). Suppose $r=0.5$ and $n$ is such that $\delta_t = 0.8$ for all $t$. LS-CS can be interpreted as KF-CS with $r=\infty$. Thus for LS-CS $a_t = 1/(1-\delta_t)=5$ always. For KF-CS, even if, at $t$, $T_t \neq T_{t-1}$, $a_t=1/(0.2+(1/5.5))=2.62$ (almost half). If $T_t$ does not change for one time instant, $a_{t+1}$ reduces to $1/(0.2+1/(a_{t}+0.5))=1.92$. If it does not change for two time instants, then $a_{t+2}$ reduces to 1.63. If $T_t$ does change and the change is a correct addition, the set $\Delta_t$ becomes smaller and so the second term of (\ref{betabound}), $\theta_t \|(x_t)_\Delta\|$, reduces. In either case, the bound reduces.

 Of course for LS-CS,  $b_t=\infty$ and so the first term of (\ref{betabound}), $\text{T1}=0$ while for KF-CS, $\text{T1} \neq 0$. But if $\sigma_{sys}^2$ and $\sigma^2$ are small and support changes slowly, $\text{T1}$ will also be small (argued earlier). When $n$ is small, the net effect is that the KF-CS bound on $\|(\beta_t)_T\|$, and hence the bound on CS-residual error, is small compared to that for LS-CS. This is the main reason that, when $n$ is very small, KF-CS error remains stable, while nothing can be said about LS-CS error. In simulations, we notice that it often becomes unstable.

\Subsection{Proof of Lemma \ref{finitedelay}}
\label{finitedelay_proof}
With  $\|w\|_\infty \le \lambda/\|A\|_1$, all results of \cite{dantzig} hold w.p. 1 (because eq 3.1  of \cite{dantzig} holds w.p. 1). From Theorem 1.1. of \cite{dantzig}, if a signal is $S$-sparse, and if $S \le \sinf$, then, the error after running the Dantzig selector is bounded by $B_*$.

The first two claims follow by induction. Consider the base case, $t=0$. The first claim holds because condition \ref{initass} of the lemma holds and because $S_r=0$ in Signal Model \ref{randomwalk_norems}. Since $|N_0| \le \smax$ and condition \ref{measmod} holds, \cite[Theorem 1.1]{dantzig} applies. Thus the second claim holds at $t=0$.
%
For the induction step, assume that the first two claims hold for $t-1$. Using the first claim for $t-1$, $|\Delta_{e,t}|=0$. Thus, $\beta_t$ is $|N_t \cup \Delta_{e,t} |= |N_t|$ sparse. Since $|N_t| \le \smax$ and condition \ref{measmod} holds, we can apply \cite[Theorem 1.1]{dantzig} to get $\|\beta_t - \betahat_t\|^2 \le B_*$. But $x_t - \xhat_{t,\CSres} = \beta_t - \betahat_t$ and so the second claim follows for $t$. By setting $\alpha = \sqrt{B_*}$ (condition \ref{thresholds}), we ensure that for any $i$ with $x_i=0$, $(\xhat_{\CSres})_i^2 = (x_i - (\xhat_{\CSres})_i)^2 \le \|x-\xhat_{\CSres}\|^2 \le B_*=\alpha^2$ (no false detects). Using this and $S_r=0$, the first claim follows for $t$.


For the third claim, it is easy to see that for any  $i \in \Delta$, if, at $t$, $(x_t)_i^2 > 2 \alpha^2 + 2B_* = 4 B_*$, then $i$ will definitely get detected at $t$. Consider a $t \in [t_j, t_{j+1}-1]$. Since $F_{j}$ holds, so at $t=t_j$, $\Delta = \add(j)$. Also, since $\alpha_{del}=0$, there cannot be false deletions and thus for any $t \in [t_j, t_{j+1}-1]$, $|\Delta| \le S_a$. Consider the worst case:  no coefficient has got detected until  $t$, i.e.  $\Delta_t=\add(j)$ and so $|\Delta_t|=S_a$. All $i \in \add(j)$ will definitely get detected at $t$ if $(x_t)_i^2 > 4B_*$ for all $i \in \add(j)$. From our model, the different coefficients are independent, and for any $i \in \add(j)$, $(x_t)_i^2 \sim \n(0, (t-t_j+1) \sigma_{sys}^2)$. Thus,
\bea
Pr((x_t)_i^2 > 4 B_*, \ \forall i \in \add(j) \ | \ F_{j} ) \se \left( 2\Q \left( \sqrt\frac{4B_*}{(t-t_{j}+1) \sigma_{sys}^2} \right) \right)^{S_a} \nn
\eea
Using the first claim, $Pr(\Nhat_t=N_t \ | \ F_{j})$ is equal to this. Thus for $t=t_j+\tau_\dett(\eps,S_a)$, $Pr(\Nhat_t=N_t \ | \ F_j) \ge 1-\eps$. Since there are no false detects; no deletions and no new additions until $t_{j+1}$, $\Nhat_t=N_t$ for $t=t_j+\tau_\dett$ implies that $E_j$ occurs. This proves the third claim.


\Subsection{Proof of  Theorem \ref{kfcs_stab}}
\label{append_kfcs}

The events $E_j$ and $F_j$ are defined in Lemma \ref{finitedelay}. At the first addition time, $t_0 = 1$, using the initialization condition, $\Nhat_{t_0-1} = N_{t_0-1}$, i.e. $F_0$ holds. Thus, by Lemma \ref{finitedelay}, $Pr(E_0) \ge 1-\eps$. Consider $t_j$ for $j>0$. Clearly \footnote{since $E_j = \{(x_{t_j+\tau_\dett})_i^2 > 4 B_*, \ \forall i \in \Delta_{t_j+\tau_\dett}\}$ and the sequence of $x_t$'s is a Markov process} $Pr(E_j|E_0,E_1, \dots E_{j-1}) = Pr(E_{j}| E_{j-1}) = Pr(E_{j}| F_{j})$. By Lemma \ref{finitedelay}, $Pr(E_j|F_j) \ge 1-\eps$.  Combining this with $Pr(E_0) \ge 1-\eps$, we get  $Pr(E_j) \ge (1-\eps)^{j+1}$ for all $j \ge 0$.


Assume that $E_j$ occurs and apply Corollary \ref{kfinitwrong_cor} with $t_* =t_j + \tau_{\dett}(\eps,S_a)$ and $t_{**}= t_{j+1}-1$. Combining the conclusion of Corollary \ref{kfinitwrong_cor} with  $Pr(E_j) \ge (1-\eps)^{j+1}$, the first claim follows.

The second and third claims follow directly from arguments in the proof of Lemma \ref{finitedelay} and $Pr(E_0 \cap E_1 \cap \dots E_{K-1}) \ge (1-\eps)^K$. %

\Subsection{Proof of Lemma \ref{kfinitwrong} and Corollary \ref{kfinitwrong_cor}}
\label{kfinitwrongproof}

Let $\xhat_{t,GAKF}$ denote the genie-aided KF (GA-KF) estimate at $t$.

Assume that the event $D$ occurs. Then, for $t > t_*$,  $\Nhat_t = N_t = N_*$, i.e. $\Delta_t := N_t \setminus \Nhat_{t-1} = N_* \setminus N_* = \phi$ (empty set) and so $\xhat_t = \xhat_{t,\text{init}}$. Let $e_t \defn x_t - \xhat_t$ and $\tilde{e}_t \defn x_t - \xhat_{t,GAKF}$.%


For simplicity of notation we assume in this proof that all variables and parameters are only along $N_*$, i.e. we let $\xhat_{t} \equiv (\xhat_{t})_{N_*}$, $e_t \equiv (e_t)_{N_*}$, $\nu_t \equiv (\nu_t)_{N_*}$, $P_{t|t-1} \equiv (P_{t|t-1})_{N_*,N_*}$, $K_{t} \equiv (K_{t})_{N_*,[1:n]}$. Let $J_t \defn I - K_t A_{N_*}$. Similarly for $\xhat_{t,GAKF}, \tilde{e}_t,  \tilde{P}_{t|t-1}, \tilde{K}_t, \tilde{J}_t$. Here $\tilde{P}_{t|t-1}, \tilde{K}_t, \tilde{J}_t$ are the corresponding matrices for GA-KF.

Let $\E[\cdot]$ denote expectation w.r.t. all random quantities conditioned on the event $D$, and let $\E[\cdot|y_1, \dots y_t]$ denote conditional expectation given $y_1, \dots y_t$ and the event $D$.


From (\ref{kftmp}), for $t > t_*$, $e_t$, $\tilde{e}_t$ and $\diff_t = e_t - \tilde{e}_t$ satisfy
\bea
e_t \se J_t e_{t-1} + J_t \nu_t - K_{t} w_t \nn \\
\label{def_e}
\tilde{e}_t \se \tilde{J}_t \tilde{e}_{t-1} + \tilde{J}_t \nu_t - \tilde{K}_{t} w_t \nn \\
\diff_{t} \se   J_t \diff_{t-1} + (J_t - \tilde{J}_t) (\tilde{e}_{t-1} + \nu_{t} ) + (\tilde{K}_{t}-K_{t}) w_t \
\label{defdt}
\eea
For $t > t_*$ both KF-CS and GA-KF run the same fixed dimensional and fixed parameter KF for $(x_t)_{N_*}$ with parameters $F \equiv I, \ Q \equiv (\sigma_{sys}^2 I_{N_*})_{N_*,N_*}, \ C \equiv A_{N_*}, \ R \equiv \sigma^2 I$, {\em but with different initial conditions}. KF-CS uses $\xhat_{t_*}$, $P_{t_*+1|t_*} \neq \E[ e_{t_*+1} e_{t_*+1}' | y_1 \dots y_{t_*}]$ while GA-KF uses the correct initial conditions, $\xhat_{t_*,GAKF}$, $\tilde{P}_{t_*+1|t_*} = \E[ \tilde{e}_{t_*+1} \tilde{e}_{t_*+1}' | y_1, \dots y_{t_*}]=\E[ \tilde{e}_{t_*+1} \tilde{e}_{t_*+1}']$. 
Since $|N_*| \le \smax$ and $\delta_{\smax} < 1$, $C \equiv A_{N_*}$ is full rank. Thus $(I, C)$ is observable. Also, since $Q$ is full rank, $(I, Q^{1/2})$ is controllable. Thus, starting from any initial condition, $P_{t+1|t}$ will converge to a  positive semi-definite, $P_*$, which is the unique solution of the discrete algebraic Riccati equation with parameters $F,Q,C,R$ \cite[Theorem 8.7.1]{ekf}.
Consequently $K_t$ and $J_t$ will also converge to  $K_* \defn P_{*} {A_{N_*}}' (A_{N_*}P_{*} {A_{N_*}}' + \sigma^2 I)^{-1}$ and $J_* \defn I - K_* A_{N_*}$ respectively. 
For $t > t_*$, the GA-KF also runs the same KF. Thus, $\tilde{P}_{t|t-1}$, $\tilde{K}_t$, $\tilde{J}_t$ will also converge to $P_*$, $K_*$, $J_*$ respectively \cite[Theorem 8.7.1]{ekf}. 
Next, we use this fact to show that the estimation errors also converge in mean square.

Using  \cite[Theorem E.5.1]{ekf}, $J_*$ is stable, i.e. its spectral radius $\rho = \rho(J_*) < 1$. Let $\eps_0 = (1- \rho)/2$. By \cite[Lemma 5.6.10]{hornjohnson}, there exists a matrix norm, denoted $\|. \|_\rho$, s.t. $\|J_{*}\|_\rho \le \rho + \eps_0 = (1+\rho)/2  < 1$.

Consider any $\eps < (1-\rho)/4$. The above results imply that there exists a $t_{\eps} > t_*$ s.t. for all $t \ge t_{\eps}$, $\|K_t - \tilde{K}_t\| < \eps$, $\|J_t - \tilde{J}_t\| < \eps$ and $\|J_{t}\|_\rho < \|J_*\|_\rho + \eps < (1+\rho)/2 + (1-\rho)/4 = (3+\rho)/4 < 1$. Now, the last set of undetected elements of $N_*$ are detected at $t_*$. Thus at $t_*$, KF-CS computes a final LS estimate, i.e. $\xhat_{t_*} ={A_{N_{*}}}^\dag y_{t_*}$,  $P_{t_*|t_*-1} = \infty$, $P_{t_*} =(A_{N_{*}}'A_{N_{*}})^{-1} \sigma^2$, $K_{t_*}=(A_{N_{*}}'A_{N_{*}})^{-1} A_{N_{*}}'$ and $J_{t_*} = 0$. None of these depend on $y_1 \dots y_{t_*}$ and hence the future values of $\xhat_t$ or of $P_{t}, J_t, K_t$ etc also do not. Hence $t_\eps$ also does not.%

Since $\tilde{P}_{t|t-1} \tends {P}_{*}$, $\tilde{P}_{t|t-1}$ is bounded. Since $\tilde{P}_t \le \tilde{P}_{t|t-1}$, $\tilde{P}_t$ is also bounded, i.e. there exists a $B < \infty$ s.t. $\trace(\tilde{P}_{t}) < B$, $\forall t$. Since $\E[\tilde{e}_t \tilde{e}_t'|y_1 \dots y_{t_*}] = \tilde{P}_{t} =   \E[\tilde{e}_t \tilde{e}_t']$, thus $\E[\|\tilde{e}_t^2\|] = \trace(\tilde{P}_{t}) < B$.


Thus, using (\ref{defdt}), the following holds for all $t \ge t_{\eps}$,
\bea
&& \E[\|\diff_t\|^2]^{1/2} \le \nn \\
&& \|M_{t,t_{\eps}}\| \ \E[\|\diff_{t_{\eps}}\|^2]^{1/2}  + \|L_{t,t_{\eps}}\| \sup_{t_{\eps} \le \tau \le t} \E[\|u_\tau\|^2]^{1/2}, \ \text{where~}   \nn \\
&&  u_{\tau} \defn (J_\tau - \tilde{J}_\tau)( \tilde{e}_{\tau-1} + \nu_\tau ) + (K_\tau - \tilde{K}_\tau) w_\tau, \nn \\
&& M_{t,t_{\eps}} \defn \prod_{k=t_{\eps}+1}^t J_k, \ \ L_{t,t_{\eps}} \defn I+ J_t + J_t J_{t-1} + .. \prod_{k=t_{\eps}+1}^t J_k  \ \ \ \ \ \ %
\label{dtbnd1}
\eea
{\em Since neither $t_\eps$, nor the matrices $J_t$ or $K_t$ for $t>t_*$, depend on $y_1, \dots y_{t_*}$, we do not need to condition the expectation on $y_1, \dots y_{t_*}$.}

Notice that 
\ben
\item  $\sup_{t_{\eps} \le \tau \le t} \E[\|u_\tau\|^2]^{1/2} \le \eps ( \sqrt{B} + \sqrt{|N_*| \sigma_{sys}^2} + \sqrt{n \sigma^2})$. 

\item $\|M_{t,t_{\eps}}\|_\rho \le \prod_{\tau=t_{\eps}+1}^t \|J_\tau\|_\rho < a^{t-t_{\eps}}$ with $a \defn (3+\rho)/4 < 1$. Thus $\|M_{t,t_{\eps}}\| \le c_{\rho,2} a^{t-t_{\eps}}$ where $c_{\rho,2}$ is the smallest real number satisfying $\|M\| \le c_{\rho,2} \|M\|_\rho$, for all size $|N_*|$  square matrices $M$ (holds because of equivalence of norms).

\item $\|L_{t,t_{\eps}}\|_\rho \le 1+a + \dots a^{t-t_{\eps}} < \frac{1}{(1-a)}$.
Thus $\|L_{t,t_{\eps}}\| \le \frac{c_{\rho,2}}{( 1 - a)}$. 
\een
Combining the above facts, for all $t \ge t_{\eps}$,
$ \E[\|\diff_t\|^2]^{1/2} \le c_{\rho,2} a^{t-t_\eps} \E[\|\diff_{t_{\eps}}\|^2]^{1/2} + C \eps$ where
$a:=(3+\rho)/4$, $C := \frac{c_{\rho,2}}{1 - a}(\sqrt{B} + \sqrt{|N_*| \sigma_{sys}^2} + \sqrt{n \sigma^2})$. Notice that $a< 1$. Consider an $\tilde{\eps} <  2C (1-\rho)/4$ and set $\eps = \tilde{\eps}/ 2C$. It is easy to see that for all $t \ge t_{\tilde{\eps}/2C} + \frac{\log (\E[\|\diff_{t_{\tilde{\eps}/ 2C}}\|^2]^{1/2} ) + \log (2 c_{\rho,2}) - \log \tilde\eps }{\log(1/a)}$, $\E[\|\diff_t\|^2]^{1/2} \le \tilde{\eps}$. Thus, conditioned on $D$, $\diff_t$ converges to zero in mean square.

By Markov's inequality, this also implies convergence in probability, i.e. for a given $\eps$, $\eps_{\err}$, there exists a $\tau_{KF}(\eps,\eps_{\err},N_*) > 0$ s.t. for all $t \ge t_* + \tau_{KF}(\eps,\eps_{\err},N_*)$,  $Pr(\|\diff_t\|^2 < \eps_{\err} \ | \ D) \ge (1-\eps)$. The proof of Corollary \ref{kfinitwrong_cor} follows directly from this.%

\bibliographystyle{IEEEbib}
\bibliography{tipnewpfmt_kfcsfullpap,tipnewpfmt} 

\begin{thebibliography}{10}

\bibitem{kfcsicip}
N.~Vaswani,
\newblock ``Kalman filtered compressed sensing,''
\newblock in {\em ICIP}, 2008.

\bibitem{kfcspap}
N.~Vaswani,
\newblock ``Analyzing least squares and kalman filtered compressed sensing,''
\newblock in {\em ICASSP}, 2009.

\bibitem{singlepixelvideo}
M.~Wakin, J.~Laska, M.~Duarte, D.~Baron, S.~Sarvotham, D.~Takhar, K.~Kelly, and
  R.~Baraniuk,
\newblock ``Compressive imaging for video representation and coding,''
\newblock in {\em Proc. Picture Coding Symposium}, April 2006.

\bibitem{just_lscs}
N.~Vaswani,
\newblock ``{LS-CS-residual (LS-CS): Compressive Sensing on Least Squares
  residual},''
\newblock {\em Accepted (with mandatory revisions) to IEEE Trans. Signal
  Processing, Arxiv preprint arXiv: 0911.5524v2}, 2010.

\bibitem{kfcsmri}
C.~Qiu, W.~Lu, and N.~Vaswani,
\newblock ``Real-time dynamic mri reconstruction using kalman filtered cs,''
\newblock in {\em ICASSP}, 2009.

\bibitem{candes}
E.~Candes, J.~Romberg, and T.~Tao,
\newblock ``Robust uncertainty principles: Exact signal reconstruction from
  highly incomplete frequency information,''
\newblock {\em IEEE Trans. Info. Th.}, vol. 52(2), pp. 489--509, February 2006.

\bibitem{donoho}
D.~Donoho,
\newblock ``Compressed sensing,''
\newblock {\em IEEE Trans. Info. Th.}, vol. 52(4), pp. 1289--1306, April 2006.

\bibitem{dantzig}
E.~Candes and T.~Tao,
\newblock ``The dantzig selector: statistical estimation when p is much larger
  than n,''
\newblock {\em Annals of Statistics}, vol. 35 (6), 2007.

\bibitem{candes_rip}
E.~Candes,
\newblock ``The restricted isometry property and its implications for
  compressed sensing,''
\newblock {\em Compte Rendus de l'Academie des Sciences, Paris, Serie I}, pp.
  589--592, 2008.

\bibitem{bpdn}
S.~S. Chen, Donoho, and M.~A. Saunders,
\newblock ``Atomic decomposition by basis pursuit,''
\newblock {\em SIAM J. Sci. Comput}, vol. 20, pp. 33--61, 1998.

\bibitem{tropp}
J.~A. Tropp,
\newblock ``Just relax: Convex programming methods for identifying sparse
  signals,''
\newblock {\em IEEE Trans. Info. Th.}, pp. 1030--1051, March 2006.

\bibitem{sparsedynamicMRI}
U.~Gamper, P.~Boesiger, and S.~Kozerke,
\newblock ``Compressed sensing in dynamic mri,''
\newblock {\em Magnetic Resonance in Medicine}, vol. 59(2), pp. 365--373,
  January 2008.

\bibitem{ekf}
T.~Kailath, A.H. Sayed, and B.~Hassibi,
\newblock {\em Linear Estimation},
\newblock Prentice Hall, 2000.

\bibitem{ldpc_cs}
S.~Sarvotham, D.~Baron, and R.~Baraniuk,
\newblock ``Compressed sensing reconstruction via belief propagation,''
\newblock in {\em Tech. rep. ECE-06-01, Dept. of ECE, Rice Univ.}, July 2006.

\bibitem{jung_etal}
H.~Jung, K.~H. Sung, K.~S. Nayak, E.~Y. Kim, and J.~C. Ye,
\newblock ``k-t focuss: a general compressed sensing framework for high
  resolution dynamic mri,''
\newblock {\em Magnetic Resonance in Medicine}, To appear.

\bibitem{bayesianCS}
S.~Ji, Y.~Xue, and L.~Carin,
\newblock ``Bayesian compressive sensing,''
\newblock {\em IEEE Trans. Sig. Proc.}, to appear.

\bibitem{schniter}
P.~Schniter, L.~Potter, and J.~Ziniel,
\newblock ``Fast bayesian matching pursuit: Model uncertainty and parameter
  estimation for sparse linear models,''
\newblock in {\em Information Theory and Applications (ITA)}, 2008.

\bibitem{giannakis}
D.~Angelosante and G.B. Giannakis,
\newblock ``Rls-weighted lasso for adaptive estimation of sparse signals,''
\newblock in {\em ICASSP}, 2009.

\bibitem{isitmodcs}
N.~Vaswani and W.~Lu,
\newblock ``Modified-cs: Modifying compressive sensing for problems with
  partially known support,''
\newblock in {\em IEEE Intl. Symp. Info. Th. (ISIT)}, 2009.

\bibitem{chenlu_tip}
C.~Qiu and N.~Vaswani,
\newblock ``Compressive sensing on the least squares and kalman filtering
  residual for real-time dynamic mri and video reconstruction,''
\newblock {\em IEEE Trans. Image Proc.}, 2009, submitted.

\bibitem{hornjohnson}
R.~Horn and C.~Johnson,
\newblock {\em Matrix Analysis},
\newblock Cambridge Univ. Press, 1985.

\bibitem{ocone}
D.~Ocone and E.~Pardoux,
\newblock ``Asymptotic stability of the optimal filter with respect to its
  initial condition,''
\newblock {\em SIAM Journal of Control and Optimization}, pp. 226--243, 1996.

\end{thebibliography}

\end{document}